\documentclass[lettersize,journal]{IEEEtran}
\usepackage{array}
\usepackage{textcomp}
\usepackage{stfloats}
\usepackage{url}
\usepackage{verbatim}
\usepackage{graphicx}
\usepackage{cite}
\usepackage{threeparttable}

\usepackage{booktabs}
\usepackage{multirow}
\usepackage{diagbox} % 加载宏包

\usepackage{hyperref}

\hypersetup{
    colorlinks=true,    % 启用彩色链接
    linkcolor=black,    % 设置链接为黑色
    citecolor=black,    % 设置引用为黑色
    filecolor=black,    % 设置文件链接为黑色
    urlcolor=black     % 设置URL链接为黑色
}

\makeatletter
\newif\if@restonecol
\makeatother

\usepackage[ruled,linesnumbered]{algorithm2e} %[ruled,vlined]{


\usepackage{subfigure}
\usepackage{subcaption}
\usepackage{booktabs}
\usepackage{ulem}
\usepackage{xcolor}
\usepackage{tikz}
\usepackage{enumerate}
\usepackage{float}
\usepackage{caption}
\usepackage{amsmath,amsthm,amssymb,amsfonts}
\usepackage{enumerate}

\usepackage[utf8]{inputenc}
\usepackage{url}
\usepackage{booktabs}
\usepackage{amssymb}
\usepackage{bbding}
\usepackage{pifont}
\usepackage{wasysym}
\usepackage{utfsym}
\usepackage{fontawesome}
\DeclareMathOperator{\Tr}{Tr}

\begin{document}
\bibliographystyle{unsrt}

\title{A Fractal-based Complex Belief Entropy for  \\ Uncertainty Measure in Complex Evidence Theory}

\author{Keming Wu, Fuyuan Xiao,~\IEEEmembership{Senior Member,~IEEE}, Yi Zhang 
        % <-this % stops a space
\thanks{Corresponding author: Fuyuan Xiao.}% <-this % stops a space
\thanks{K. Wu, F. Xiao  and Y. Zhang are with the School of Big Data and Software Engineering, Chongqing University 401331, China (e-mail: xiaofuyuan@cqu.edu.cn)}}

\markboth{Journal of \LaTeX\ Class Files,~Vol.~14, No.~8, August~2021}%
{Shell \MakeLowercase{\textit{et al.}}: A Sample Article Using IEEEtran.cls for IEEE Journals}

\noindent\footnotesize\textbf{Notice:} \\
This work has been submitted to the IEEE for possible publication. \\
Copyright may be transferred without notice, after which this version may no longer be accessible.

\IEEEpubid{0000--0000/00\$00.00~\copyright~2021 IEEE}
\makeatletter
\renewcommand{\maketag@@@}[1]{\hbox{\m@th\normalsize\normalfont#1}}%
\makeatother

\maketitle

\begin{abstract}
Complex Evidence Theory (CET), an extension of the traditional D-S evidence theory, has garnered academic interest for its capacity to articulate uncertainty through Complex Basic Belief Assignment (CBBA) and to perform uncertainty reasoning using complex combination rules. Nonetheless, quantifying uncertainty within CET remains a subject of ongoing research. To enhance decision-making, a method for Complex Pignistic Belief Transformation (CPBT) has been introduced, which allocates CBBAs of multi-element focal elements to subsets. CPBT's core lies in the fractal-inspired redistribution of the complex mass function. This paper presents an experimental simulation and analysis of CPBT's generation process along the temporal dimension, rooted in fractal theory. Subsequently, a novel Fractal-Based Complex Belief (FCB) entropy is proposed to gauge the uncertainty of CBBA. The properties of FCB entropy are examined, and its efficacy is demonstrated through various numerical examples and practical application.

% Complex evidence theory, as a generalized D-S evidence theory, has attracted academic attention because it can well express uncertainty by means of complex basic belief assignment (CBBA), and realize uncertainty reasoning by complex combination rule. However, the uncertainty measurement in complex evidence theory is still an open issue. In order to make better decisions, a complex pignistic belief transformation (CPBT) method has been proposed to assign CBBAs of multi-element focal elements to subsets. The essence of CPBT is the redistribution of complex mass function by means of the concept of fractal. In this paper, based on fractal theory, experimental  simulation and analysis have been carried out on the generation process of CPBT in time dimension. Then, a new fractal-based complex belief (FCB) entropy is proposed to measure the uncertainty of CBBA. Finally, the properties of FCB entropy are analyzed, and several examples are used to verify its effectiveness.
\end{abstract}

\begin{IEEEkeywords}
Complex Evidence Theory, Complex Basic Belief Assignment, Complex Pignistic Belief Transformation, Uncertainty Measurement, Fractal-based Complex Belief Entropy.
\end{IEEEkeywords}

\section{Introduction}
% \IEEEPARstart
% {F}{ractals} are ubiquitous in nature and were initially introduced to quantify the length of coastlines. The field of fractal theory, as it stands today, is a highly regarded and dynamic area of study. As interest in fractal theory grows, its applications have extended across a variety of disciplines. The essence of fractal theory lies in examining objective phenomena through the lens of fractal dimension, offering novel insights for analyzing complex information systems \cite{Qiang2022fractal}. Moreover, the self-similarity characteristic of fractals holds promise for its utilization in information analysis.

% {F}{ractal} widely exists in nature and was first proposed to measure the length of the coast. The fractal theory established today is a very popular and active new theory and discipline. With people's attention to fractal theory, it has been widely used in many different fields. The most basic feature of fractal theory is to study objective things with fractal dimension, which provides a new perspective for the analysis of complex information systems \cite{Qiang2022fractal}. In addition, the self-similarity of fractal also has the prospect of application in information analysis.

{P}{recise} decision-making necessitates the quantification of uncertainty, dynamic and reliability analysis, a critical aspect within information analysis theory \cite{zhou2019survey, wang2022expert, liu2022orientational, meng2023novel, liao2023asynchronous, wang2024complex, chen2024risk}. To address this, various theories have been introduced, including the extended Z-number \cite{zhu2022zacm}, D number \cite{FrameworkDNT2023}, rough sets \cite{Fujita2020Hypotheses,ye2021novel}, random permutation sets \cite{deng2022RPS}, Dempster-Shafer evidence theory (DSET) \cite{dempster1967upper, shafer1976mathematical} and so on \cite{cao2022network}. DSET, in particular, and its extensions \cite{fei2022optimization, fei2024evidential} are favored for their nuanced handling of uncertain information probabilities and have been extensively applied across numerous domains.
These applications span decision-making \cite{garg2020multiattribute,zhou2023large}, multi-source information fusion \cite{Xiao2022GEJS}, classification \cite{huang2023higherR, huang2023fractal, zhang2024divergence}, reasoning \cite{xu2020evidence,tang2019perturbation}, and diagnosis \cite{fu2022extended, xu2024fault}, among others \cite{cao2020interpretability, li2022reliability,ni2021towards}.

% Because of the need for accurate decision-making, uncertainty measurement is an essential issue in information analysis theory. In order to solve this problem, many relevant theories have been proposed, such as Z-number \cite{zhu2022zacm}, intuitionistic fuzzy set \cite{tao2021dynamic,song2019divergencebased}, rough set \cite{Fujita2020Hypotheses,ye2021novel}, random permutation set \cite{deng2022RPS}, Dempster-Shafer evidence theory (DSET), etc \cite{deng2020information,xiong2021conflicting,chang2021transparent}. Owing to its advantages in probability distribution of uncertain information, DSET and its branches \cite{hua2022consensus} are widely used in many fields, such as decision-making \cite{che2022maximum,Xiao2022Generalizeddivergence,liao2021process,garg2020multiattribute,zhou2022consensus,zhou2023large}, pattern recognition \cite{deng2021improved,liu2020evidence}, multi-source information fusion 
%  \cite{li2021multisource,Xiao2022GEJS,zhu2022divergence}, classification decision \cite{Shang2021compound,Xiao2022Acomplexweighted}, reasoning \cite{fu2021evidential,xu2020evidence,wang2022fusion}, fault diagnosis \cite{meng2022fault}, etc \cite{Chenxingyuan2022SREM,9117190,wei2021velocity,li2022reliability,ni2021towards,CHEN2021104438}.

% In probability theory, Shannon entropy is usually used to represent the information volume, which solves the measurement problem of information quantification. However, 
% In probability theory, the probability distribution of different events is directly used as the basis for decision-making. 
DSET serves as an extension of traditional probability theory, wherein probabilities are assigned to the power set of elements to encapsulate the represented information \cite{xiong2021conflicting, chang2021transparent}. This is achieved through the use of a mass function, which quantifies the probability of each power set, known as the basic belief assignment (BBA). Grappling with the concepts of understanding and quantifying uncertainty lies at the heart of DSET \cite{liu2021new, zhang2022bsc}. BBA in DSET is influenced not only by the inherent probability of an event but also by the uncertainty associated with the allocation of sets. However, the mass function in DSET operates within the real number field, presenting limitations when capturing the nuances of data fluctuations at specific stages. DSET falls short in managing uncertain information in the complex number field \cite{QuantumBPA2023}. Consequently, the aforementioned methods struggle to accurately represent information within this domain. To address this, Xiao introduced Complex Evidence Theory (CET), also known as Generalized DSET \cite{xiao2020generalization,xiao2020generalized}. Unlike DSET, CET's Complex Basic Belief Assignment (CBBA) effectively encapsulates information in the complex field, and its complex evidence combination rules are adept at handling uncertainty reasoning with complex numbers \cite{zhang2024Gaussian,yang2023exponential}. Notably, CET has demonstrated significant potential and advantages for information modeling in quantum fields \cite{Xiao2022NQMF,xiao2023generalized, Xiao2023QuantumXentropy, wu2024novel, he2024quantumrule}.
In essence, CET offers a novel framework for uncertain information reasoning, with the uncertainty measure of CBBA being one of its most critical aspects.

To be specific, the uncertainty in DSET is typically composed of discord and non-specificity. Discord reflects the conflicts among different elements within the framework \cite{LEI2022112136}, while non-specificity captures the uncertainty that arises during the distribution of BBA \cite{zhu2024fractal}. Currently, two primary approaches are used to quantify BBA uncertainty. The first approach leverages entropy-based methods, drawing on concepts such as Shannon entropy. This category includes various entropy measures like Hohle entropy \cite{hohle1982entropy}, weighted Hartley entropy \cite{dubois1987properties}, Pal et al.’s entropy \cite{pal1992uncertainty}, Jousselme et al.’s entropy \cite{jousselme2006measuring}, Deng’s entropy \cite{Deng2020ScienceChina}, Jirou{\v{s}}ek and Shenoy's entropy \cite{jirouvsek2018new}, Cui and Deng's Plausibility entropy \cite{PlEntropy2023}, among others \cite{cao2019extraction,liu2021double}. In contrast, the non-entropy approach sidesteps the potential divergence between DSET and probability theory by directly defining uncertainty measures on the framework using belief and likelihood functions. Examples include Song and Wang's uncertainty measure \cite{wang2018uncertainty}, Yang and Han's distance-based total uncertainty measures \cite{yang2016new,han2018belief}, and the total uncertainty measure for interval-valued belief structures \cite{yager2008entropy}.

\IEEEpubidadjcol
While the DSET and its extensions have been instrumental in advancing the field of uncertainty quantification, the existing methods for uncertainty measurement within CET are still in their nascent stages. Most of the research efforts in CET have been building upon the foundational work established by classical DSET. This includes the exploration of the uncertainty measures that are compatible with the complex nature of CBBAs, which is a departure from the traditional real-number-based BBAs in DSET. The existing complex-valued Deng entropy \cite{pan2023complex} transforms Deng entropy into complex form, and quantum belief entropy \cite{wu2024novel} uses density matrix to measure the uncertainty of CBBA. 
Although these methods use complex numbers for uncertainty modeling, they do not fully utilize the influence of the intersection of different focal elements and maximize the use of complex number representation information.

To address limitations in uncertainty measurements like complex-valued Deng entropy and quantum belief entropy within CET, fractal theory offers a compelling approach. Fractals, known for their self-similarity and ability to analyze complex systems, have found applications across various disciplines. The fractal dimension provides new insights into complex information systems, aligning with CET's need for advanced uncertainty quantification. Drawing inspiration from \cite{zhou2022fractal}, a novel fractal-based Complex Pignistic Belief Transformation (CPBT) process has been introduced, leading to the development of Fractal-Based Complex Belief (FCB) entropy. This new metric, FCB entropy, is designed to capture the overall uncertainty in CBBAs more effectively. Theoretical analysis and practical applications substantiate the viability of FCB entropy, showcasing its potential to enhance decision-making and system understanding within CET by leveraging the full capacity of complex numbers.

The remainder of the paper is structured as follows. In Section \ref{se2}, some basic concepts of DSET and CET are briefly reviewed, and then fractal theory and some common entropies are briefly introduced. In Section \ref{se3}, a CPBT generation process based on fractal idea is proposed firstly, and then FCB entropy is proposed based on the above process. Then the properties of FCB entropy are analyzed and proved in Section \ref{se4}. Then numerical example and a algorithm with its practical application are used to verify the effectiveness of FCB entropy in Section \ref{se5}. Section \ref{se6} is focused on conclusions and future prospects.
	
\section{Preliminaries}\label{se2}
How to model and measure the uncertainty and dynamics of the system \cite{wang2018exploiting, wang2017onymity, Wang2020CommunicatingSA, wang2022modelling} have attracted a lot of attentions. In this section, the basic theories of DSET and CET are briefly reviewed, and then fractal theory and some common entropies in evidence theory are introduced.

\subsection{Dempster-Shafer evidence theory}
\newtheorem{definition}{Definition}
\newtheorem{exa}{Example}
\newtheorem{pro}{Property}
\newtheorem{axi}{Axiom}
\newtheorem{lem}{Lemma}
\newtheorem{cas}{Case}
\newtheorem{proo}{Proof}

\begin{definition}(Frame of Discernment).
	\rm Let a set be exhaustible and the elements in it are mutually exclusive, which can be denoted as $\Theta =\{{{e}_{1}},{{e}_{2}},\cdots ,{{e}_{n}}\}$.
	Then the set $\Theta $ is called a frame of discernment (FoD). The power set of $\Theta $ is denoted as ${{2}^{\Theta }}=\{\varnothing ,\{{{e}_{1}}\},\{{{e}_{2}}\},\cdots \{{{e}_{1}},{{e}_{2}}\},\cdots ,\Theta \}$, where $\varnothing $ is an empty set \cite{dempster1967upper, shafer1976mathematical}.
\end{definition}

\begin{definition}(Mass Function).
\rm Let $m$ be a mss fuction in FoD $\Theta =\{{{e}_{1}},{{e}_{2}},\cdots ,{{e}_{n}}\}$, which is a mapping:$m:{{2}^{\Theta }}\to \mathbb{R}$ \cite{dempster1967upper, shafer1976mathematical} satisfying 
\begin{equation}
% \small
m(\varnothing )=0,
\end{equation}
\begin{equation}
% \small
\sum\limits_{{{A}_{i}}\in {{2}^{\Theta }}}{m({{A}_{i}})}=1.
\end{equation}
If $m({{A}_{i}})>0$, then ${{A}_{i}}$ is called a focal element.
\end{definition}

% \begin{definition}
% 	(Pignistic Probability Transformation).
% 	\rm	Given a FoD $\Theta =\{{{e}_{1}},{{e}_{2}},\cdots ,{{e}_{n}}\}$ composed of $n$ elements and the corresponding BBA $m$, its pignistic probability transformation (PPT) is defined as \cite{smets2005decision}:
% 	\begin{equation}
%         \small
% 		BetP({{e}_{i}})=\sum\limits_{{{e}_{i}}\in {{A}_{i}}and{{A}_{i}}\in {{2}^{\Theta }}}{\frac{m({{A}_{i}})}{\left| {{A}_{i}} \right|}},
% 	\end{equation}
% where $\left| {{A}_{i}} \right|$ represents the cardinality of ${{A}_{i}}$.
% \end{definition}

\subsection{Complex evidence theory}
DSET is an important evidential reasoning theory for dealing with uncertain information. It not only allocates the probability to a single element, but also to a set of multiple elements, which contains more information than the traditional probability distribution. Complex evidence theory is a generalization of DSET and has the ability to process complex information.
\begin{definition}(Complex Mass Function).
	\rm	Let $\mathbb{M}$ be a complex mass function (CMF) in FoD $\Theta =\{{{e}_{1}},{{e}_{2}},\cdots ,{{e}_{n}}\}$, which is a mapping: $\mathbb{M}:{{2}^{\Theta }}\to \mathbb{C}$ satisfying \cite{xiao2020generalization,xiao2020generalized}
	\begin{equation}
        % \small
		\mathbb{M}(\varnothing )=0,
	\end{equation}
	\begin{equation}
        % \small
	\mathbb{M}({{A}_{i}})=\mathbf{m}({{A}_{i}}){{e}^{i\theta ({{A}_{i}})}},{{A}_{i}}\in {{2}^{\Theta }},
\end{equation}
which satisfies
	\begin{equation}
        % \small
		\sum\limits_{{{A}_{i}}\in {{2}^{\Theta }}}{\mathbb{M}({{A}_{i}})}=1,
	\end{equation}
 where $\sqrt{i}=-1$ and $\mathbf{m}({{A}_{i}})\in [0,1]$ is denoted as the magnitude of CMF $\mathbb{M}({{A}_{i}})$. $\theta ({{A}_{i}})\in [-2\pi ,2\pi ]$ is denoted as the phase of CMF $\mathbb{M}({{A}_{i}})$.
 Through Euler formula, CMF can be converted as follows:
 
	\begin{equation}\label{module}
        % \small
	\mathbb{M}({{A}_{i}})=x+yi,{{A}_{i}}\in {{2}^{\Theta }},
	\end{equation}
	\begin{equation}
        % \small
	\left| \mathbb{M}({{A}_{i}}) \right|=\mathbf{m}({{A}_{i}})=\sqrt{{{x}^{2}}+{{y}^{2}}}.
\end{equation}
% And the phase of CMF $\mathbb{M}({{A}_{i}})$ is defined by

% \begin{equation}
%         \small
% 	\theta \left( A \right)=\left\{ 
% 	\begin{matrix}
% 		\arctan \left( \frac{y}{x} \right), & x>0  \\
% 		\frac{\pi }{2}, & x=0,y>0  \\
% 		-\frac{\pi }{2}, & x=0,y<0  \\
% 		\arctan \left( \frac{y}{x} \right)+\pi , & x<0,y\ge 0  \\
% 		\arctan \left( \frac{y}{x} \right)-\pi , & x<0,y<0 
% 	\end{matrix} 
% \right..
% \end{equation}
If $\left| \mathbb{M}({{A}_{i}}) \right|>0$, ${{A}_{i}}$ is called a focal element in CMF.
\end{definition}

%Definition 2.5(Dempster’s Rule of Combination)
	\begin{definition} (Commitment Degree).
	\rm	The commitment degree $\mathbb{C}om({{A}_{i}})$ represents the support degree for ${{A}_{i}}$, which is expressed as \cite{xiao2020generalization,xiao2020generalized}
		\begin{equation}
  % \small
\mathbb{C}om({{A}_{i}})=\frac{\left| \mathbb{M}({{A}_{i}}) \right|}{\sum\nolimits_{{{B}_{i}}\in {{2}^{\Theta }}}{\left| \mathbb{M}({{B}_{i}}) \right|}}.
		\end{equation}

	\end{definition}
	
%	Definition 2.6(Quantum Probability)
	\begin{definition}(Complex Pignistic Belief Transformation).
	\rm Complex pignistic belief transformation (CPBT) is used to assign CBBAs of multi-element focal elements to subsets, which is defined as \cite{Xiao2022NQMF}
	 \begin{equation}\label{eq:10}
  % \small
	CBet({{A}_{i}})=\sum\limits_{{{A}_{i}}\subseteq {{B}_{i}}\in {{2}^{\Theta }}}{\frac{\mathbb{M}({{B}_{i}})}{\left| {{B}_{i}} \right|}},
	 \end{equation}
 where $\left| {{B}_{i}} \right|$ represents the cardinality of ${{B}_{i}}$ and $\left| {{A}_{i}}\cap {{B}_{i}} \right|$ represents the cardinality of the intersection of ${{A}_{i}}$ and ${{B}_{i}}$.
	\end{definition}

\begin{definition}(Interference Effect).
\rm	Interference effect refers to the residual term generated when calculating the added complex modulus, which is defined as \cite{huang2023some}
	\begin{equation}
        % \small
		\begin{aligned}
	\underset{B\subseteq \Theta }{\mathop{IE(B)}}\,
	&=\sum\limits_{{{A}_{i}}\subseteq B,{{A}_{j}}\subseteq B,{{A}_{i}}\ne {{A}_{j}}}{2\mathbf{m}({{A}_{i}})\mathbf{m}({{A}_{j}})\cos (\theta ({{A}_{i}})-\theta ({{A}_{j}}))}\\
	&={{\left| \sum\limits_{{{A}_{i}}\subseteq B}{\mathbb{M}({{A}_{i}})} \right|}^{2}}-\sum\limits_{{{A}_{i}}\subseteq B}{{{\mathbf{m}}^{2}}({{A}_{i}})},
		\end{aligned}
	\end{equation}
\end{definition}
\begin{definition}(CBBA Exponential Negation).
	\rm Given a FoD $\Theta =\{{{e}_{1}}$, ${{e}_{2}}$, $\cdots $, ${{ e }_{n}}\}$ and the power set of $\Theta $ is ${{2}^{\Theta }}=\{\varnothing ,\{{{e}_{1}}\},\{{{e}_{2}}\},\cdots ,\{{{e}_{1}},{{e}_{2}}\},\cdots ,\Theta \}$. Let a CBBA $\mathbb{M}$ be defined on ${{2}^{\Theta }}$.
	Then the CBBA exponential negation is defined by \cite{yang2023exponential}
	\begin{equation}
        % \small
	\overline{\mathbb{M}}({{A}_{i}})=\frac{\sum\limits_{B\ne \Theta ,{{B}_{i}}\in {{2}^{\Theta }}}{{{e}^{-\mathbb{M}({{B}_{i}})\cdot \left| {{A}_{i}}\cap \overline{{{B}_{i}}} \right|}}}}{\sum\limits_{{{A}_{i}}\ne \varnothing ,{{A}_{i}}\in {{2}^{\Theta }}}{\sum\limits_{{{B}_{i}}\ne \Theta ,{{B}_{i}}\in {{2}^{\Theta }}}{{{e}^{-\mathbb{M}({{B}_{i}})\cdot \left| {{A}_{i}}\cap \overline{{{B}_{i}}} \right|}}}}},A\ne \varnothing,
	\end{equation}
% and
% \begin{equation}
%         \small
% 	\overline{\mathbb{M}}(\varnothing )=0,
% \end{equation}
where $\overline{{{B}_{i}}}=\Theta -{{B}_{i}}$ and $\left| {{A}_{i}}\cap \overline{{{B}_{i}}} \right|$ represent the cardinality of set ${{A}_{i}}\cap \overline{{{B}_{i}}}$.
% Another form of the formula can be expressed as
% \begin{equation}
% \small
% \overline{\mathbb{M}}({{A}_{i}})=\frac{\sum\limits_{B\ne \Theta ,{{B}_{i}}\in {{2}^{\Theta }}}{{{e}^{-\mathbb{M}({{B}_{i}})\cdot \left| {{A}_{i}}\cap \overline{{{B}_{i}}} \right|}}}}{\sum\limits_{{{B}_{i}}\ne \Theta ,{{B}_{i}}\in {{2}^{\Theta }}}{{{e}^{-\mathbb{M}({{B}_{i}})\cdot \left| \overline{{{B}_{i}}} \right|\cdot {{2}^{n-1}}}}}},A\ne \varnothing,
% \end{equation}
\end{definition}

\begin{definition}\label{combine}
\rm	Let ${\mathbb{M}_{1}}$ and ${\mathbb{M}_{2}}$ are two CBBAs defined in the FOD $\Omega $. The complex evidence combination rule is defined as \cite{xiao2020generalization,xiao2020generalized}
	\begin{equation}\label{De3.1}
	\mathbb{M}(C)=\left\{
	\begin{aligned}
		&	\frac{1}{1-K}\sum\limits_{A\cap B=C}{{\mathbb{M}_{1}}(A){\mathbb{M}_{2}}(B)},&C\ne \varnothing ,  \\
		&	0,&C=\varnothing,   \\
		\end{aligned}
	 \right.
	\end{equation}
with
\begin{equation}
	\mathbb{K}=\sum\limits_{A\cap B=\varnothing }{{\mathbb{M}_{1}}(A){\mathbb{M}_{2}}(B)},
\end{equation}
where $\mathbb{K}$ is the conflict coefficient between CBBAs ${\mathbb{M}_{1}}$ and ${\mathbb{M}_{2}}$.
\end{definition}

\subsection{Entropy}
The concept of entropy originated from the physical thermodynamic system and gradually extended to the field of information to measure the amount of information. Some commonly used entropies are reviewed below.
\begin{definition}(Shannon Entropy).
	\rm Given the probability distribution of all events in a framework $P=\{{{p}_{1}},\cdots ,{{p}_{n}}\}$, then the Shannon entropy is defined as below \cite{shannon1948mathematical}:
	\begin{equation}
 % \small
	{{E}_{s}}(P)=-\sum\limits_{i=1}^{n}{{{p}_{i}}\log {{p}_{i}}}.
	\end{equation}
	% if and only if ${{p}_{1}}={{p}_{2}}=\cdots ={{p}_{n}}=\frac{1}{n}$, ${{E}_{s}}(P)$ gets the maximum value which can be calculated as
	% \begin{equation}
 % \small
	% {{E}_{s}}(P)=-\sum\limits_{i=1}^{n}{\frac{1}{n}\log \frac{1}{n}}=\log n.
	% \end{equation}
\end{definition}
	\begin{definition}(Weighted Hartley Entropy).
		\rm Weighted Hartley entropy is extended from Hartley information formula, which is defined as follows \cite{higashi1982measures}:
\begin{equation}
% \small
	{{E}_{H}}(m)=\sum\limits_{{{A}_{i}}\in {{2}^{\Theta }}}{m({{A}_{i}})\log \left| {{A}_{i}} \right|}.
\end{equation}
	\end{definition}
\begin{definition}(Pal et al.’s Entropy).
	\rm Given a mass function $m$ in an $n$-element FoD $\Theta =\{{{e}_{1}},{{e}_{2}},\cdots ,{{ e }_{n}}\}$, then the Pal et al.’s entropy is defined as below \cite{pal1992uncertainty}:
		\begin{equation}
  % \small
		{{E}_{p}}(m)=-\sum\limits_{{{A}_{i}}\in {{2}^{\Theta }}}{m({{A}_{i}})\log (\frac{m({{A}_{i}})}{\left| {{A}_{i}} \right|})}.
	\end{equation}
% if and only if $m({{A}_{i}})=\frac{\left| {{A}_{i}} \right|}{\sum\limits_{{{A}_{i}}\in {{2}^{\Theta }}}{\left| {{A}_{i}} \right|}}$, ${{E}_{p}}(m)$ gets the maximum value which can be calculated as

% 	\begin{equation}
%  \small
%  \begin{aligned}
% {{E}_{p}}(m)&=-\sum\limits_{{{A}_{i}}\in {{2}^{\Theta }}}{\frac{\left| {{A}_{i}} \right|}{\sum\limits_{{{A}_{i}}\in {{2}^{\Theta }}}{\left| {{A}_{i}} \right|}}\log (\frac{1}{\sum\limits_{{{A}_{i}}\in {{2}^{\Theta }}}{\left| {{A}_{i}} \right|}})} \\
%       &=\log (\left| \Theta  \right|\cdot {{2}^{\left| \Theta  \right|-1}}).
%  \end{aligned}
% \end{equation}

\end{definition}
\begin{definition}(Deng Entropy).
	\rm Given a mass function $m$ in an $n$-element FoD $\Theta =\{{{e}_{1}},{{e}_{2}},\cdots ,{{ e }_{n}}\}$, then the Deng entropy is defined as below \cite{Deng2020ScienceChina}:
	\begin{equation}
 % \small
	{{E}_{d}}(m)=-\sum\limits_{{{A}_{i}}\in {{2}^{\Theta }}}{m({{A}_{i}})\log \frac{m({{A}_{i}})}{{{2}^{\left| {{A}_{i}} \right|}}-1}},
	\end{equation}
% if and only if $m({{A}_{i}})=\frac{{{2}^{\left| {{A}_{i}} \right|}}-1}{\sum\limits_{{{A}_{i}}\in {{2}^{\Theta }}}{{{2}^{\left| {{A}_{i}} \right|}}-1}}$, ${{E}_{d}}(m)$ gets the maximum value which can be calculated as
% \begin{equation}
% \small
% \begin{aligned}
%     {{E}_{d}}(m)&=-\sum\limits_{{{A}_{i}}\in {{2}^{\Theta }}}{\frac{{{2}^{\left| {{A}_{i}} \right|}}-1}{\sum\limits_{{{A}_{i}}\in {{2}^{\Theta }}}{{{2}^{\left| {{A}_{i}} \right|}}-1}}\log \frac{1}{\sum\limits_{{{A}_{i}}\in {{2}^{\Theta }}}{{{2}^{\left| {{A}_{i}} \right|}}-1}}}\\
%     &=\log ({{3}^{\left| \Theta  \right|}}-{{2}^{\left| \Theta  \right|}}).
% \end{aligned}
% \end{equation}

\end{definition}
\begin{definition}(Zhou et al.s’ Entropy).
	\rm Given a mass function $m$ in an $n$-element FoD $\Theta =\{{{e}_{1}},{{e}_{2}},\cdots ,{{ e }_{n}}\}$, then the Zhou et al.s’ entropy is defined as below \cite{zhou2020weight}:
	\begin{equation}
 % \small
		{{E}_{z}}(m)=-\sum\limits_{{{A}_{i}}\in {{2}^{\Theta }}}{m({{A}_{i}})\log (\frac{m({{A}_{i}})}{{{2}^{\left| A \right|}}-1}{{e}^{\frac{\left| A \right|-1}{\left| X \right|}}})}.
	\end{equation}
\end{definition}
\begin{definition}
	(Cui et al.s’ Entropy).
\rm 	Given a mass function $m$ in an $n$-element FoD $\Theta =\{{{e}_{1}},{{e}_{2}},\cdots ,{{ e }_{n}}\}$, then the Cui et al.s’ entropy is defined as below \cite{cui2019improved}:
\begin{equation}
% \small
	{{E}_{c}}(m)=-\sum\limits_{{{A}_{i}}\in {{2}^{\Theta }}}{m({{A}_{i}})\log }\left( \frac{m({{A}_{i}})}{{{2}^{\left| {{A}_{i}} \right|}}-1}{{e}^{\sum\limits_{\begin{smallmatrix} 
					{{B}_{i}}\in {{2}^{\Theta }} \\ 
					{{B}_{i}}\ne {{A}_{i}} 
			\end{smallmatrix}}{\frac{\left| {{A}_{i}}\cap {{B}_{i}} \right|}{{{2}^{\left| \Theta  \right|}}-1}}}} \right).
\end{equation}
\end{definition}
\begin{definition}(FB Entropy).
	\rm Given a mass function $m$ in an $n$-element FoD $\Theta =\{{{e}_{1}},{{e}_{2}},\cdots ,{{ e }_{n}}\}$, then the FB entropy is defined as below \cite{zhou2022fractal}:
	\begin{equation}
 % \small
		{{E}_{FB}}(m)=-\sum\limits_{{{A}_{i}}\in {{2}^{\Theta }}}{{{m}_{F}}({{A}_{i}})\log {{m}_{F}}({{A}_{i}})},
	\end{equation}
where ${{m}_{F}}({{A}_{i}})$ is mass function after fractal. And ${{m}_{F}}({{A}_{i}})$ is defined as:
\begin{equation}
% \small
	{{m}_{F}}({{A}_{i}})=\frac{m({{A}_{i}})}{{{2}^{\left| {{A}_{i}} \right|}}-1}+\sum\limits_{{{A}_{i}}\subset {{A}_{j}}}{\frac{m({{A}_{j}})}{{{2}^{\left| {{A}_{j}} \right|}}-1}},
\end{equation}
	% if and only if ${{m}_{F}}({{A}_{i}})=\frac{1}{{{2}^{\left| {{A}_{i}} \right|}}-1}$ or $m(\Theta )=1$, ${E_{FB}}(m)$ gets the maximum value which can be calculated as
	% \begin{equation}
 % \small
	% 	{{E}_{FB}}(m)=\log ({{2}^{\left| \Theta  \right|}}-1).
	% \end{equation}
\end{definition}
% \begin{definition}\label{def2.16}
% 	(Generalized Entropy with its Maximum).
% 	\rm Because the entropies mentioned above are defined in the real number field, while CBBA is defined in the complex number field. The above entropies cannot be directly used in CET. A generalized entropy in CET is defined as follows:
% 	\begin{equation}
%  \small
% 	\mathbb{E}(\mathbb{M})=-\sum\limits_{{{A}_{i}}\in {{2}^{\Theta }}}{\mathbb{C}om({{A}_{i}})\cdot \log (\frac{\mathbb{C}om({{A}_{i}})}{h({{A}_{i}})})}.
% 	\end{equation}
% when the generalized entropy is used in BBA, it can be expressed as
% \begin{equation}
% \small
% 	E(m)=-\sum\limits_{{{A}_{i}}\in {{2}^{\Theta }}}{m({{A}_{i}})\cdot \log (\frac{m({{A}_{i}})}{h({{A}_{i}})})}.
% \end{equation}
% \end{definition}

% 新添加的定义 阳的论文里的定义，记得response时说明
\begin{definition}\label{def2.16}
	(Generalized entropy with its maximum).
	\rm Because the entropies mentioned above are defined in the real number field, while CBBA is defined in the complex number field. The above entropies cannot be directly used in CET. A generalized entropy in CET is defined as follows \cite{yang2023exponential}:
	\begin{equation}
	\mathbb{E}(\mathbb{M})=-\sum\limits_{{{A}_{i}}\in {{2}^{\Theta }}}{\mathbb{C}om({{A}_{i}})\cdot \log (\frac{\mathbb{C}om({{A}_{i}})}{h({{A}_{i}})})}.
	\end{equation}
when the generalized entropy is used in BBA, it can be expressed as
\begin{equation}
	E(m)=-\sum\limits_{{{A}_{i}}\in {{2}^{\Theta }}}{m({{A}_{i}})\cdot \log (\frac{m({{A}_{i}})}{h({{A}_{i}})})}.
\end{equation}
\end{definition}

\begin{definition}\label{def2.17}
(Complex-valued Deng entropy).
\rm	Given a CBBA $\mathbb{CM}$ defined on a FOD $\mathbb{E}$, then complex-valued Deng entropy is defined as follows \cite{pan2023complex}:
\begin{equation}
	\begin{aligned}
	    C{{E}_{d}} &=\left\| -\sum\limits_{A\subseteq \mathbb{E}}{\left\| \mathbb{CM}\left( A \right) \right\|\ln \frac{\mathbb{CM}\left( A \right)}{{{2}^{\left| A \right|}}-1}} \right\| \\
     & =\left\| -\sum\limits_{A\subseteq \mathbb{E}}{\left\| m\left( A \right){{e}^{i\pi {{\theta }_{A}}}} \right\|\ln \frac{m\left( A \right){{e}^{i\pi {{\theta }_{A}}}}}{{{2}^{\left| A \right|}}-1}} \right\| \\
     & =\left\| -\sum\limits_{A\subseteq \mathbb{E}}{m\left( A \right)\ln \frac{m\left( A \right){{e}^{i\pi {{\theta }_{A}}}}}{{{2}^{\left| A \right|}}-1}} \right\|
	\end{aligned}
\end{equation}
 \end{definition}

% 新添加的定义用于对比
\begin{definition}\label{def2.18}
(Quantum belief entropy)
\rm	Let a CBBA $\mathbb{M}$ define on a FOD $\Theta =\left\{ {{\theta }_{1}},{{\theta }_{2}},\cdots ,{{\theta }_{n}} \right\}$ and the associated density matrix is $\rho $. Then QB entropy of CBBA is defined as follows \cite{wu2024novel}:
\begin{equation}\label{eq35}
	{{E}_{Q}}(\mathbb{M})=S(\rho )+\sum\limits_{i\ne j}{Q{{I}_{ij}}},
\end{equation}
	\begin{equation}
S( \rho  )=\sum\limits_{{{\theta }_{k}}\in \Theta }{\left| P\left( {{\theta }_{k}} \right)\log_{2}P({{\theta }_{k}})\right|},
	\end{equation}
where $P( {{\theta }_{k}} )=\Tr( {{\mathbb{M}}_{{{\theta }_{k}}}}^{+}{{\mathbb{M}}_{{{\theta }_{k}}}}\rho  )$. 
 \end{definition}

\section{Fractal-based complex belief entropy}\label{se3}
% In this section, based on the idea of fractal, a generation process of CPBT is proposed, which can well reveal the distribution of CBBAs of multi-element focal elements. Then FCB entropy is proposed based on the generation process of CPBT, and the maximum entropy model of FCB is deduced.

% \subsection{Theoretical analysis of CPBT from fractal perspective}
% Compared with the traditional probability theory, DSET and CET can express more information because the mass function is assigned to all power sets in FoD. However, in the real world, the probability of a certain event is usually expressed by the probability theory of Bayesian distribution. Therefore, transforming the mass function of evidence theory into probability is a crucial issue. In DSET, PPT is widely used in practical problems as an important method to convert BBA into probability. In \cite{zhou2022fractal}, a fractal-based PPT generation process is proposed, which provides a new perspective for describing the relationship between BBA and probability. In CET, the research on the relationship between CBBA and probability has not been widely carried out. CPBT is a common method to convert CBBAs of multi-element focal elements into singletons. Then $\mathbb{C}om$ is used to play the role of decision instead of probability in CET \cite{xiao2021complex}. Inspired by \cite{zhou2022fractal}, this paper proposes a fractal generation process of CPBT in order to fill the gap in the research of this direction in CET. 

In this section, the importance of CET in uncertainty modeling is discussed in detail at first. Then we introduce a fractal-inspired generation process for Complex Pignistic Belief Transformation (CPBT) that effectively elucidates the distribution of CBBAs with multi-element focal elements. Subsequently, we propose Fractal-Based (FCB) entropy, which is grounded in the CPBT generation process, and deduce the maximum entropy model for FCB.

\subsection{The importance of Deng entropy and CET in modeling uncertainty}
The classical Dempster-Shafer (D-S) evidence theory, grounded in the real number field, adeptly addresses general decision-making problems \cite{chen2023novel}. Deng entropy, as an important uncertainty measurement method in D-S evidence theory, has attracted more and more attention \cite{zhao2024linearity, pan2023complex} and has been improved from multiple perspectives \cite{cui2019improved, tang2023improved, kharazmi2023deng} and applied in different fields \cite{ozkan2023new}. Deng entropy takes the effect of cardinality of focal elements in BBA into account in the measurement of uncertainty, which is an important theoretical basis for our FCB entropy \cite{Qiang2022fractal}.

However, the intricacies of real-world scenarios necessitate a more comprehensive representation of information. The use of complex numbers captures the nuances that real numbers alone might not fully express. Traditional D-S evidence theory's representation is confined to real space, which may fall short in effectively capturing and processing uncertain information, especially when dealing with data that exhibit phase angle information. CBBA offers a more sophisticated and holistic approach to encapsulate all pertinent information about a BBA. Furthermore, while it is true that CBBAs need to be transformed into probabilities for decision-making purposes, this transformation is not straightforward due to the complex nature of the information. The complex mass function must be converted into a form that can be compared and aggregated, which is inherently possible with probabilities. The transformation process itself benefits from the richness of the CBBA, as it provides a more detailed and comprehensive basis for the decision-making process.

% In the context of CBBA, uncertainty is characterized by a complex number with an amplitude component and a phase component. The amplitude reflects the extent of the evidence body's support for the proposition, akin to traditional BBA, while the phase angle incorporates information regarding the reliability of this support. The significance of CBBA extends beyond these interpretations, with vast potential for further exploration and application. 

% D-S evidence theory encounters limitations when attempting to encapsulate the uncertainty inherent in complex information. In contrast, Complex Evidence Theory (CET) excels at unearthing the uncertainty within CBBA in the complex domain, offering significant advantages in navigating complex situations. 

Scholarly work on CET's application in uncertainty modeling is extensive. A negation method for CBBA in CET has been introduced and applied in multi-source information fusion \cite{Xiao2022NQMF, yang2023exponential}. Additionally, a  complex Gaussian fuzzy numbers-based uncertainty modeling technique within CET  has facilitated the generation of CBBA, enhancing the precision of post-fusion decision-making beyond traditional approaches \cite{zhang2024Gaussian}. A negation-based multisource information fusion outperforms well-known related approaches with higher classification accuracy and robustness \cite{Xiao2022NQMF}.
 
\subsection{Theoretical Analysis of CPBT from a Fractal Perspective}
DSET and CET offer richer information representation compared to traditional probability theory, as they assign mass functions to the entire power set in the Frame of Discernment (FoD). However, in practical scenarios, the probability of specific events is often articulated through Bayesian probability theory. Thus, the transformation of evidence theory's mass function into probability is of pivotal importance. In DSET, PPT is frequently employed to convert BBA into probability. \cite{zhou2022fractal} introduces a fractal-based PPT generation process, offering a novel lens to understand the connection between BBA and probability. In CET, the translation of CBBA into probability remains relatively unexplored. CPBT is a standard approach for transforming multi-element focal CBBAs into singularities. The decision-making role in CET is then assumed by $\mathbb{C}om$, which replaces the function of probability \cite{Xiao2022NQMF}. Motivated by \cite{zhou2022fractal}, this paper presents a fractal-based CPBT generation process to address the research gap in this area within CET.

For the 2-element Frame of Discernment (FoD)  $X=\{{{x}_{1}},{{x}_{2}}\}$, the Complex Pignistic Belief Transformation (CPBT) generation process can be conceptualized as the gradual allocation of a two-element CBBA to a single element over the course of 
$n$ iterations. As  $n$ approaches infinity, the CBBA associated with the two elements progressively diminishes to zero, while the CBBA of singletons stabilizes, with the information content ultimately converging to a constant value. A detailed example is provided below to elucidate the CPBT generation process.
\begin{figure}[ht] %H为当前位置，!htb为忽略美学标准，htbp为浮动图形
	\centering %图片居中
	\includegraphics[width=0.5\textwidth]{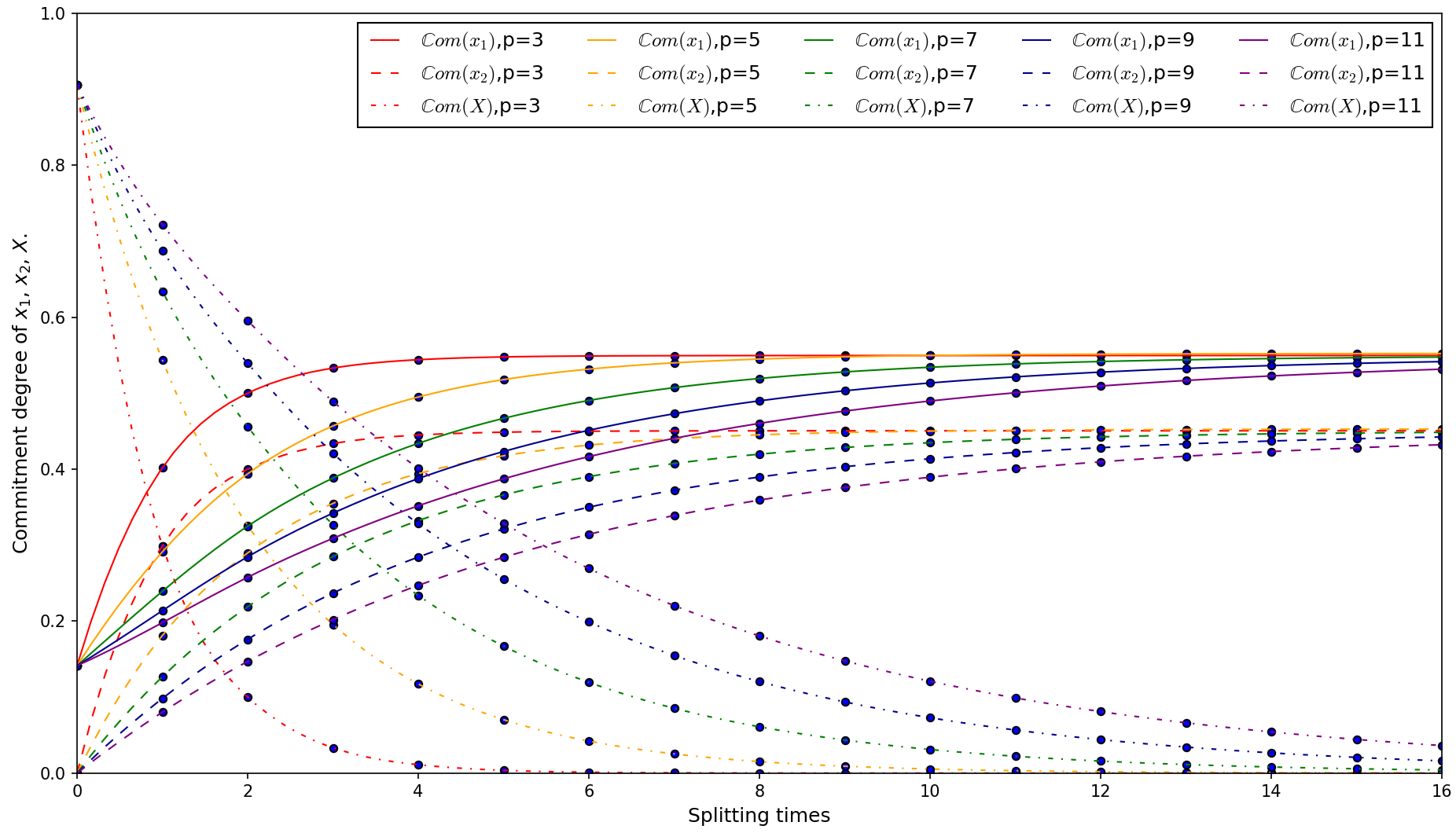} %插入图片，[]中设置图片大小，{}中是图片文件名
	\caption{Regardless of the value of $p$, the final distribution of $\mathbb{C}om({{x}_{1}})$, $\mathbb{C}om({{x}_{2}})$ and $\mathbb{C}om({{X}})$ is stable and tends to the same value.} %最终文档中希望显示的图片标题
	\label{Fig.2} %用于文内引用的标签
\end{figure}
\begin{exa}
\label{exa:1}
\rm	Given a FoD $X=\{{{x}_{1}},{{x}_{2}}\}$, and the CBBA defined in it is shown as below:
	\begin{displaymath}
		\begin{aligned}
	&\mathbb{M}({{x}_{1}})=0.1414{{e}^{i\arctan (-1.0000)}}=0.1-0.1i,\\
	&	\mathbb{M}({{x}_{1}},{{x}_{2}})=0.9055{{e}^{i\arctan (0.1111)}}=0.9+0.1i.
		\end{aligned}
	\end{displaymath}
The CBBA after $n$ times of splitting is shown as follows:
\begin{displaymath}
	\begin{aligned}
&{{\mathbb{M}}^{n}}({{x}_{1}})={{\mathbb{M}}^{n-1}}({{x}_{1}})+\frac{1}{p}{{\mathbb{M}}^{n-1}}({{x}_{1}},{{x}_{2}}),\\
&{{\mathbb{M}}^{n}}({{x}_{2}})={{\mathbb{M}}^{n-1}}({{x}_{2}})+\frac{1}{p}{{\mathbb{M}}^{n-1}}({{x}_{1}},{{x}_{2}}),\\
&{{\mathbb{M}}^{n}}({{x}_{1}},{{x}_{2}})=(1-\frac{2}{p}){{\mathbb{M}}^{n-1}}({{x}_{1}},{{x}_{2}}),
	\end{aligned}
\end{displaymath}
where $p\ge 3$, ${{\mathbb{M}}^{n}}({{x}_{i}})$ means $n$ times splitting of $\mathbb{M}({{x}_{i}})$. $p$ represents the size of $\mathbb{M}({{x}_{1}},{{x}_{2}})$ allocated to $\mathbb{M}({{x}_{1}})$ and $\mathbb{M}({{x}_{2}})$, which can also be understood as the allocation speed of $\mathbb{M}({{x}_{1}},{{x}_{2}})$ per unit time. When $p$ takes different values, the CBBA after n iterations are different. For the iterated CBBA,  $\mathbb{C}om({{A}_{i}})$ is used to show its support for the element ${{A}_{i}}$, similar to BBA. The results are shown in Figure \ref{Fig.2}.

It is not difficult to understand the proposed CPBT generation process from Example \ref{exa:1}. For FoD with $n$ elements, give a CBBA $\mathbb{M}$. The production process of CPBT based on fractal is that the CBBA value of multi-element focal elements are evenly distributed to single element focal elements in a certain proportion in the process of time change, gradually reducing the influence of non-specificity until finally disappearing. Finally, the CBBA of Bayesian distribution is obtained.
\end{exa}

\begin{figure}[ht] %H为当前位置，!htb为忽略美学标准，htbp为浮动图形
	\centering %图片居中
	\includegraphics[width=0.35\textwidth]{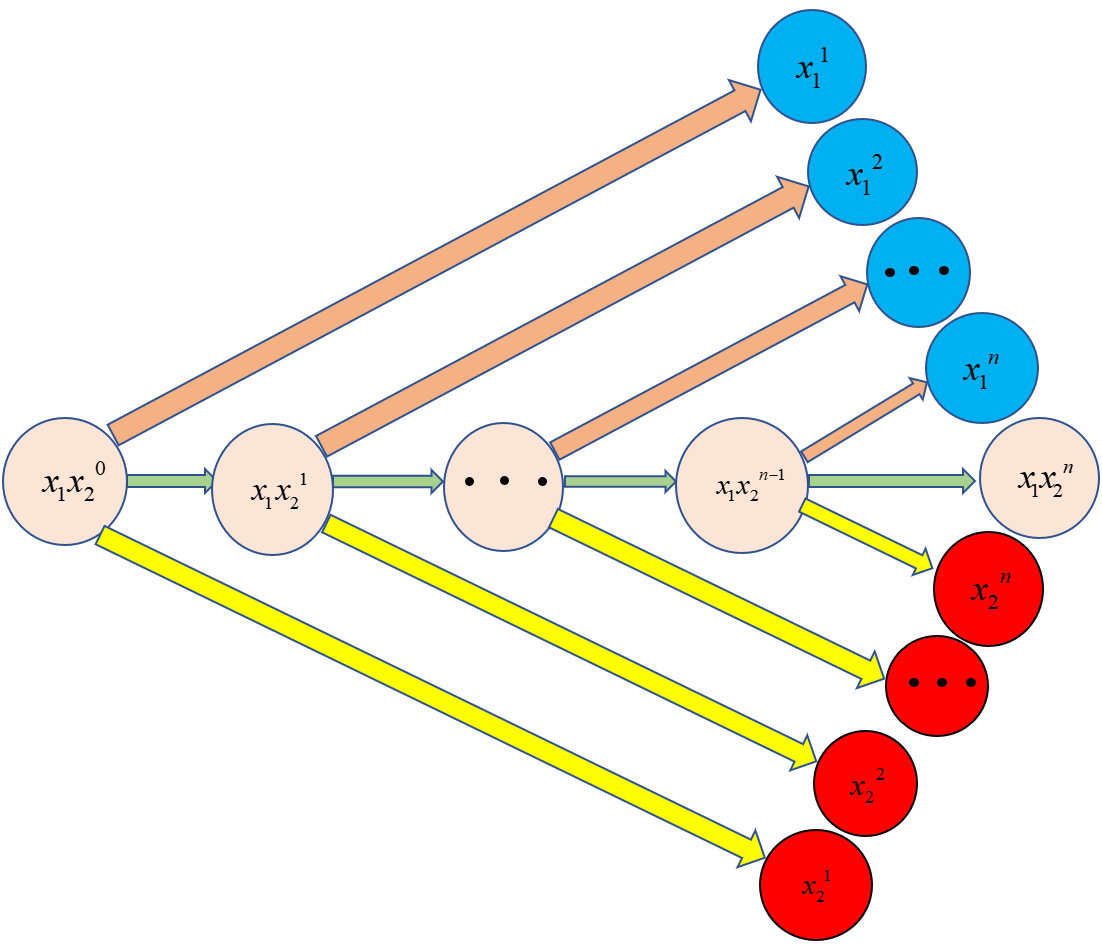} %插入图片，[]中设置图片大小，{}中是图片文件名
	\caption{The process of $x_{1}$, $x_{2}$ allocation to singletons in Example 1 is self-similar. } %最终文档中希望显示的图片标题
	\label{Fig.18} %用于文内引用的标签
\end{figure}

A hallmark characteristic of fractals is their self-similarity, and the proposed CPBT generation process embodies this through a self-similar iterative approach. The fractal diagram illustrating Example 1 is depicted in Figure \ref{Fig.18}.

\subsection{Fractal-based complex belief entropy}
Drawing on the aforementioned CPBT generation process, we introduce a novel measure of uncertainty for CBBA known as Fractal-Based Complex Belief (FCB) entropy. As demonstrated in the two examples, varying parameter values result in different extents of allocation to multi-element focal elements. To enhance the representation of allocation rationality within FCB entropy, it is stipulated that the focal element's mass is uniformly distributed among its constituent elements.

% \begin{figure}[H] %H为当前位置，!htb为忽略美学标准，htbp为浮动图形
% 	\centering %图片居中
% 	\includegraphics[width=0.35\textwidth]{Fig.17} %插入图片，[]中设置图片大小，{}中是图片文件名
% 	\caption{Flow chart of FCB entropy’s measurement. } %最终文档中希望显示的图片标题
% 	\label{Fig.17} %用于文内引用的标签
% \end{figure}
\begin{definition}(FCBBA).\label{def3.1}
	\rm Given a FoD $\Theta =\{{{e}_{1}},{{e}_{2}},\cdots ,{{ e }_{n}}\}$ and its corresponding CBBA $\mathbb{M}$, For any ${{A}_{k}}$ belonging to ${{2}^{\Theta }}$, its CBBA after fractal is defined by
		\begin{equation}\label{eq17}
        % \small
		{{\mathbb{M}}_{F}}({{A}_{k}})=\frac{\mathbb{M}({{A}_{k}})}{{{2}^{\left| {{A}_{k}} \right|}}-1}+\sum\limits_{{{A}_{k}}\subseteq {{B}_{k}}\cap \left| {{A}_{k}} \right|<\left| {{B}_{k}} \right|}{\frac{\mathbb{M}({{B}_{k}})}{{{2}^{\left| {{B}_{k}} \right|}}-1}}.
	\end{equation}
The new set $2_{FB}^{\Theta }$ composed of ${{\mathbb{M}}_{F}}({{A}_{k}})=\frac{\mathbb{M}({{A}_{k}})}{{{2}^{\left| {{A}_{k}} \right|}}-1}+\sum\limits_{{{A}_{k}}\subseteq {{B}_{k}}\cap \left| {{A}_{k}} \right|<\left| {{B}_{k}} \right|}{\frac{\mathbb{M}({{B}_{k}})}{{{2}^{\left| {{B}_{k}} \right|}}-1}}$ is called  fractal-based complex basic belief assignment (FCBBA).
\end{definition}

\begin{definition}\label{def3.2}
	(FCB entropy).
	\rm Given a FoD $\Theta =\{{{e}_{1}},{{e}_{2}},\cdots ,{{ e }_{n}}\}$ and its corresponding FCBBA ${{\mathbb{M}}_{F}}({{A}_{k}})=\frac{\mathbb{M}({{A}_{k}})}{{{2}^{\left| {{A}_{k}} \right|}}-1}+\sum\limits_{{{A}_{k}}\subseteq {{B}_{k}}\cap \left| {{A}_{k}} \right|<\left| {{B}_{k}} \right|}{\frac{\mathbb{M}({{B}_{k}})}{{{2}^{\left| {{B}_{k}} \right|}}-1}}$, then the FCB entropy is defined by:
	\begin{equation}\label{eq.31}
 % \small
		\begin{aligned}
					{{\mathbb{E}}_{FCB}}(\mathbb{M})
					&=-\sum\limits_{{{A}_{k}}\in {{2}^{\Theta }}}{\mathbb{C}o{{m}_{F}}({{A}_{k}})\log (\mathbb{C}o{{m}_{F}}({{A}_{k}}))},\\
		\end{aligned}
	\end{equation}
where $\mathbb{C}o{{m}_{F}}({{A}_{k}})$ represents the support degree for ${{A}_{k}}$ in $2_{FB}^{\Theta }$ and $\mathbb{C}o{{m}_{F}}({{A}_{k}})$ in the FCB entropy is defined by
\begin{equation}\label{eq.32}
% \small
	\mathbb{C}o{{m}_{F}}({{A}_{k}})=\frac{\left| {{\mathbb{M}}_{F}}({{A}_{k}}) \right|}{\sum\limits_{B\in \Theta }{\left| {{\mathbb{M}}_{F}}({{B}_{k}}) \right|}},
\end{equation}
where $\sum\limits_{B\in \Theta }{\left| {{\mathbb{M}}_{F}}({{B}_{k}}) \right|}$ is used to normalize $\mathbb{C}o{{m}_{F}}({{A}_{k}})$, which ensures that the value of $\mathbb{C}o{{m}_{F}}({{A}_{k}})$ is between [0,1]. So $\mathbb{C}o{{m}_{F}}({{A}_{k}})$ also satisfies the following equation
\begin{equation}\label{eq.34}
% \small
	\sum\limits_{{{A}_{k}}\in {{2}^{\Theta }}}{\mathbb{C}o{{m}_{F}}({{A}_{k}})}=1.
\end{equation}
When substituting (\ref{eq.32}) into (\ref{eq.31}), (\ref{eq.31}) can be transformed into another form,
\begin{equation}
% \small
{{\mathbb{E}}_{FCB}}(\mathbb{M})=-\sum\limits_{{{A}_{k}}\in {{2}^{\Theta }}}{\frac{\left| {{\mathbb{M}}_{F}}({{A}_{k}}) \right|}{\sum\limits_{B\in \Theta }{\left| {{\mathbb{M}}_{F}}({{B}_{k}}) \right|}}\log \left( \frac{\left| {{\mathbb{M}}_{F}}({{A}_{k}}) \right|}{\sum\limits_{B\in \Theta }{\left| {{\mathbb{M}}_{F}}({{B}_{k}}) \right|}} \right)}.
\end{equation}
The formula for calculating the modulus of the focal element in  $\mathbb{C}o{{m}_{F}}({{A}_{k}})$ is defined as follows:
\begin{equation}
% \small
	\begin{aligned}
\left| {{\mathbb{M}}_{F}}({{A}_{k}}) \right|
&=\left| \frac{\mathbb{M}({{A}_{k}})}{{{2}^{\left| {{A}_{k}} \right|}}-1}+\sum\limits_{{{A}_{k}}\subseteq {{B}_{k}}\cap \left| {{A}_{k}} \right|<\left| {{B}_{k}} \right|}{\frac{\mathbb{M}({{B}_{k}})}{{{2}^{\left| {{B}_{k}} \right|}}-1}} \right|\\
&=\sqrt{{{\sum\limits_{{{C}_{k}}\in {{B}_{k}}}{\left| \frac{\mathbb{M}({{C}_{k}})}{{{2}^{\left| {{C}_{k}} \right|}}-1} \right|}}^{2}}+\sum\limits_{{{A}_{k}}\subseteq {{B}_{k}}\cap \left| {{A}_{k}} \right|<\left| {{B}_{k}} \right|}{IE({{B}_{k}})}}
	\end{aligned}
\end{equation}
where $IE({B}_{k})$ is the interference function in CET and in FCB entropy it is defined by
\begin{equation}
% \small
	\begin{aligned}
\underset{{B}_{k}\subseteq \Theta }{\mathop{IE({B}_{k})}}\,
&=\sum\limits_{{{X}_{i}}\subseteq {{B}_{k}},{{X}_{j}}\subseteq {{B}_{k}},{{X}_{i}}\ne {{X}_{j}}}{2\mathbf{m}({{X}_{i}})\mathbf{m}({{X}_{j}})\cos (\theta ({{X}_{i}})-\theta ({{X}_{j}}))}\\
&={{\left| \sum\limits_{{{X}_{i}}\subseteq {B}_{k}}{\mathbb{M}({{X}_{i}})} \right|}^{2}}-\sum\limits_{{{X}_{i}}\subseteq {B}_{k}}{{{\mathbf{m}}^{2}}({{X}_{i}})}.
	\end{aligned}
\end{equation}
$\mathbb{M}({{X}_{i}})$ and $\mathbb{M}({{X}_{j}})$ have the same form as CBBA in CET
\begin{equation}
% \small
	\mathbb{M}({{X}_{i}})=u+vi.
\end{equation}
$\theta ({{X}_{i}})$ can be calculated by the following formula
\begin{equation}
% \small
	\theta \left( {{X}_{i}} \right)=\left\{ 
	\begin{matrix}
		\arctan \left( \frac{v}{u} \right), & u>0  \\
		\frac{\pi }{2}, & u=0,v>0  \\
		-\frac{\pi }{2}, & u=0,v<0  \\
		\arctan \left( \frac{v}{u} \right)+\pi , & u<0,v\ge 0  \\
		\arctan \left( \frac{v}{u} \right)-\pi , & u<0,v<0  \\
	\end{matrix}
 \right.
\end{equation}
After Euler transformation, $\mathbb{M}({{X}_{i}})$ and $\mathbb{M}({{X}_{j}})$ in $IE({B}_{k})$ can be expressed as
\begin{equation}
% \small
	\mathbb{M}({{X}_{i}})=\mathbf{m}({{X}_{i}}){{e}^{i\theta ({{X}_{i}})}},
\end{equation}
\begin{equation}
% \small
	\mathbb{M}({{X}_{j}})=\mathbf{m}({{X}_{j}}){{e}^{i\theta ({{X}_{j}})}},
\end{equation}
In the complex plane, the vector sum of $\mathbb{M}({{X}_{i}})$ and $\mathbb{M}({{X}_{j}})$ can be expressed by 
\begin{equation}
% \small
	\mathbb{M}({{X}_{i}})+\mathbb{M}({{X}_{j}})=\mathbf{m}({{X}_{i}}){{e}^{i\theta ({{X}_{i}})}}+\mathbf{m}({{X}_{j}}){{e}^{i\theta ({{X}_{j}})}}.
\end{equation}
It can be calculated from cosine theorem that
\begin{equation}
% \small
\begin{aligned}
&{{\left| \mathbf{m}({{X}_{i}}){{e}^{i\theta ({{X}_{i}})}}+\mathbf{m}({{X}_{j}}){{e}^{i\theta ({{X}_{j}})}} \right|}^{2}}\\
&={{\left| \mathbf{m}({{X}_{i}}){{e}^{i\theta ({{X}_{i}})}} \right|}^{2}} +{{\left| \mathbf{m}({{X}_{j}}){{e}^{i\theta ({{X}_{j}})}} \right|}^{2}}+IE(X),
\end{aligned}
\end{equation}
where $IE(X)=2\mathbf{m}({{X}_{i}})\mathbf{m}({{X}_{j}})\cos (\theta ({{X}_{i}})-\theta ({{X}_{j}}))$.

\end{definition}
\begin{axi}\label{axi3.1}
\rm	When CBBA degenerates to BBA, FCB entropy degenerates to FB entropy, that is, In DSET, FCB entropy and FB entropy are equivalent. The proof \ref{pro3.1} is in the appendix.
\end{axi}

\subsection{Maximum fractal-based complex belief entropy}
The maximum entropy principle is a criterion for selecting the distribution of random variables that best conforms to the objective situation, also known as the maximum information principle. Generally, only one distribution has the maximum entropy. Choosing this distribution with maximum entropy as the distribution of the random variable is an effective criterion for decision analysis. Therefore, the derivation of the maximum FCB entropy model is very important. The definition of maximum FCB entropy is as follows.
\begin{definition}(Maximum FCB entropy).\label{Def3.3}
	\rm Given a FoD $\Theta =\{{{e}_{1}},{{e}_{2}},\cdots ,{{ e }_{n}}\}$ and the CBBA $\mathbb{M}$ in it, the maximum FCB entropy is 
	\begin{equation}
 % \small
		\mathbb{E}_{FCB}^{\max }(\mathbb{M})=\log ({{2}^{\left| \Theta  \right|}}-1)=\log ({{2}^{n}}-1),
	\end{equation}
when $\mathbb{C}o{{m}_{F}}({{A}_{k}})=\frac{1}{{{2}^{\left| \Theta  \right|}}-1}$. The proof \ref{pro3.3} is in the appendix.

FCB entropy represents an extension of Shannon entropy, with its maximum entropy model paralleling that of Shannon's while also being capable of addressing decision-making issues in the real world. As an advancement over FB entropy, the maximum entropy model of FCB entropy retains the physical interpretation of FB entropy and offers additional advantages.
\end{definition}

\section{The properties of the proposed FCB entropy}\label{se4}
For the total uncertainty measurement of BBA, different methods generally have 10 properties \cite{abellan2008requirements}, which are often used for the analysis of uncertain methods. For the analysis of FCB entropy, the 10 properties have been extended to CET to measure the feasibility and applicability of FCB entropy. The properties of some uncertain methods are summarized in Table \ref{tab1}.
\begin{table*}
	\centering
	\caption{Summary of properties of some entropies.}\label{tab1}%添加标题 设置标签
	% \resizebox{.8\textwidth}{!}{
		\begin{threeparttable}
			\begin{tabular}{lccccc}
				\toprule
				& \multicolumn{5}{c}{Properties} \\ \cmidrule(lr){2-6}
				Measures &  Probabilistic consistency & Set consistency & Additivity &Subadditivity&Maximum entropy \\
				\midrule
				Hartley Entropy \cite{higashi1982measures}&\XSolidBrush&\Checkmark &\Checkmark&\Checkmark&\Checkmark\\
				%		Confusion Measurement \cite{höhle1982entropy}&√&×&×&$A_{2}$\\
				%		Discord Measurement \cite{klir1990uncertainty}&√&×&√&$A_{2}$\\
				Pal et al.’s entropy \cite{pal1992uncertainty}&\Checkmark&\Checkmark&\Checkmark&\XSolidBrush&\XSolidBrush\\
				Zhou et al.’s measure \cite{zhou2020weight}&\Checkmark&\XSolidBrush&\Checkmark&\XSolidBrush&\Checkmark\\
				Deng entropy \cite{Deng2020ScienceChina}&\Checkmark&\XSolidBrush&\XSolidBrush&\XSolidBrush&\Checkmark\\
				JS entropy \cite{jirouvsek2018new}&\Checkmark&\Checkmark&\Checkmark&\XSolidBrush&\Checkmark\\
            QB entropy \cite{wu2024novel}&\Checkmark&\XSolidBrush&\XSolidBrush&\XSolidBrush&\XSolidBrush\\
				FB entropy \cite{zhou2022fractal}&\Checkmark&\XSolidBrush&\Checkmark&\Checkmark&\Checkmark\\
				FCB entropy&\Checkmark&\XSolidBrush&\Checkmark&\Checkmark&\Checkmark\\
				%		Yang and Han’s method \cite{yang2016new}&×&×&√&$A_{1}$\\
				%		QB entropy&√&×&×&$A_{5}$\\
				\bottomrule
			\end{tabular}
		\end{threeparttable}
	% }
\end{table*}
\begin{pro}[\textit{Probabilistic consistency}]\label{pro:1}
\rm	When all focal elements are singletons, the distribution of mass function is similar to Bayesian distribution, and the total uncertainty measurement should be degenerated to Shannon entropy.

For FCB entropy, $\forall \left| {{A}_{i}} \right|>1,\mathbb{M}({{A}_{i}})=0$,$\left| {{\mathbb{M}}_{F}}({{A}_{i}}) \right|=\left| \frac{\mathbb{M}({{A}_{i}})}{{{2}^{\left| {{A}_{i}} \right|}}-1} \right|$.

\noindent \textbf{Case 1.}
when CBBA degenerates BBA and according to Axiom \ref{axi3.1}, FCB entropy is degenerated to FB entropy. And the BBA in FoD $\Theta =\{{{e}_{1}},{{e}_{2}},\cdots ,{{ e }_{n}}\}$ satisfies $\sum\limits_{i=1}^{n}{m({{e}_{i}})}=1$, substitute it into (\ref{eq.32}):
\begin{equation}
% \small
\begin{aligned}
{{\mathbb{E}}_{FCB}}(m)&=-\sum\limits_{i=1}^{n}{m({{e}_{i}})\log (m({{e}_{i}}))}\\
&={{E}_{FB}}(m)={{E}_{s}}(m). 
\end{aligned}
\end{equation}
It is obvious that it satisfies Property \ref{pro:1}.

\noindent \textbf{Case 2.}
When FCB entropy is expressed in the form of CBBA, its probability consistency cannot be judged in the usual form. Expand it to use $\mathbb{C}om({{A}_{i}})$ to express the probability of focal element ${{A}_{i}}$. The CBBA in FoD $\Theta =\{{{e}_{1}},{{e}_{2}},\cdots ,{{ e }_{n}}\}$ satisfies $\sum\limits_{i=1}^{n}{\mathbb{M}({{e}_{i}})}=1$ and $\sum\limits_{i=1}^{n}{\mathbb{C}o{{m}_{F}}({{e}_{i}})}=1$, then substitute it into (\ref{eq.32}):

\begin{equation}
% \small
	\begin{aligned}
	{{\mathbb{E}}_{FCB}}(\mathbb{M})
	&=-\sum\limits_{i=1}^{n}{\mathbb{C}o{{m}_{F}}({{e}_{i}})\log (\mathbb{C}o{{m}_{F}}({{e}_{i}}))}\\
	&=-\sum\limits_{i=1}^{n}{\mathbb{C}om({{e}_{i}})\log (\mathbb{C}om({{e}_{i}}))}.
	\end{aligned}
\end{equation}
In CET, $\mathbb{C}om({{e}_{i}})$ represents the degree of support for a focal element ${{e}_{i}}$, which is similar to the role of probability in probability theory. The simplified FCB entropy has the same form as Shannon entropy, and is similar to the information it expresses. The generalized probability consistency of FCB entropy in the complex number field can also be satisfied.

%Property 2. Set Consistency
\begin{pro}[\textit{Set  correlation}]\label{pro:2}
%	\rm In DSET, when $m(A)=1,A\ne \varnothing $, the total uncertainty should become the Hartley measure $\log \left| A \right|$.
	\rm The number of elements in the focal element will also affect the information modeling when the complex mass function is allocated. The traditional set consistency is controversial, but it provides a new perspective for the study of the set-related properties of FCB entropy.
	
	\noindent \textbf{Explanation:} Hartley's information formula measures the amount of information in probability theory, while FCB entropy measures the uncertainty in FoD. In the case of the same dimension, evidence theory can express more information than probability theory. For a FoD with $n$ elements, the framework of evidence theory contains ${{2}^{n}}$ focal elements, while probability theory only contains $n$ elements. Therefore, it is reasonable that the maximum value of FCB entropy is greater than the maximum value of Shannon entropy. For some commonly used entropies, their maximum entropies are summarized in Table \ref{tab2}. The maximum value of different entropy changes with the number of elements as shown in Figure \ref{Fig.1}.
\end{pro}

\begin{figure}[H] %H为当前位置，!htb为忽略美学标准，htbp为浮动图形
	\centering %图片居中
	\includegraphics[width=0.5\textwidth]{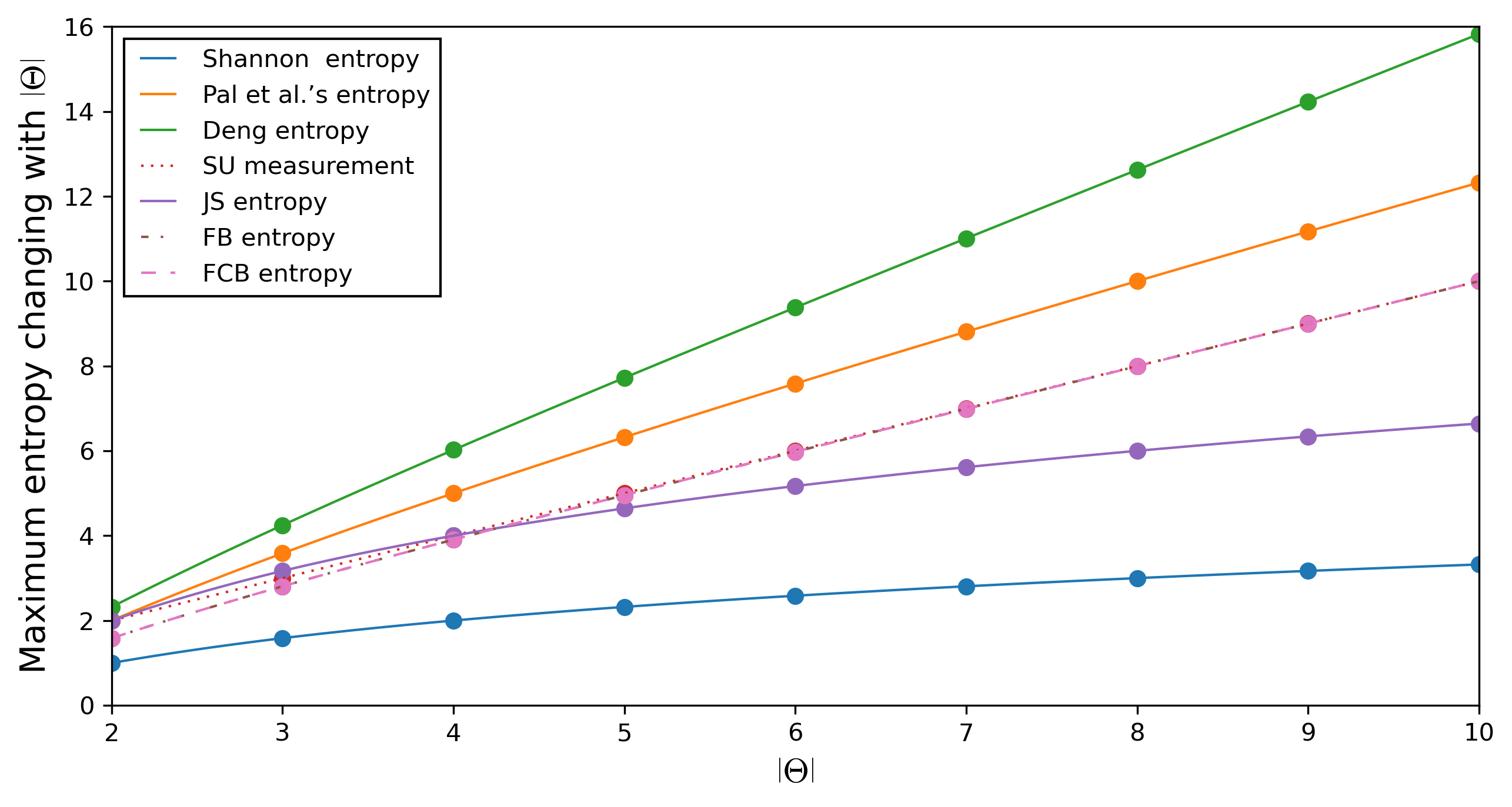} %插入图片，[]中设置图片大小，{}中是图片文件名
	\caption{As the number of elements in FoD increases, the maximum value of most entropies is greater than the maximum value of Shannon entropy.} %最终文档中希望显示的图片标题
	\label{Fig.1} %用于文内引用的标签
\end{figure}
\end{pro}
\begin{table*}
	\centering
	\caption{Summary of maximum of some entropies.}\label{tab2}%添加标题 设置标签
	% \resizebox{.5\textwidth}{!}{
		\begin{threeparttable}
			\begin{tabular}{lllll}
				\toprule
				& \multicolumn{4}{c}{Methods} \\ \cmidrule(lr){2-5}
				& Shannon entropy \cite{shannon1948mathematical} & Pal et al.’s entropy \cite{pal1992uncertainty}& Deng entropy \cite{Deng2020ScienceChina}&SU measurement \\
%				\midrule
				Maximum &$\log \left( \left| \Theta  \right| \right)$&$\log \left( \left| \Theta  \right|\cdot {{2}^{\left| \Theta  \right|-1}} \right)$&$\log \left( {{3}^{\left| \Theta  \right|}}-{{2}^{\left| \Theta  \right|}} \right)$&$\left| \Theta  \right|$\\
				\midrule
				& JS entropy \cite{jirouvsek2018new} & FB entropy \cite{zhou2022fractal} &FCB entropy\\
%				\midrule
				Maximum&$2\log \left( \left| \Theta  \right| \right)$&$\log ({{2}^{\left| \Theta  \right|}}-1)$&$\log ({{2}^{\left| \Theta  \right|}}-1)$& \\
				%		Yang and Han’s method \cite{yang2016new}&×&×&√&$A_{1}$\\
				%		QB entropy&√&×&×&$A_{5}$\\
				\bottomrule
			\end{tabular}
		\end{threeparttable}
	% }
\end{table*}
%Property 3. Range
\begin{pro}[\textit{Range}]\label{pro.3}
	\rm Given a FoD $\Theta =\{{{e}_{1}},{{e}_{2}},\cdots ,{{ e }_{n}}\}$, the range of the total uncertainty measurements should be in $[0,\log \left| \Theta  \right|]$.
% \begin{proo}

\begin{table*}[!htbp]
	\centering
	% \LARGE
	%	\setlength{\tabcolsep}{15pt}
	\caption{Results of each iteration of FCB entropy.}\label{tab:5}%添加标题 设置标签
	% \resizebox{.5\textwidth}{!}{
		\begin{threeparttable}
			\begin{tabular}{llllllllllll}
				\toprule
				& \multicolumn{11}{c}{Times} \\ \cmidrule(lr){2-12}
				Measures& 0& 1 &  2 & 3 & 4&5&6&7&8&9&10 \\
				\midrule
				FCB entropy&1.3458&1.4633&1.4734&1.4834&1.4834&1.4834&1.4834&1.4834&1.4834&1.4834&1.4834\\
				\bottomrule
			\end{tabular}
		\end{threeparttable}
	% }
\end{table*}

\noindent \textbf{Proof 3.}
\rm	When there exists a elemnt in the FoD satisfying $\mathbb{M}({{e}_{i}})=1$, the minimum of FCB entropy is $\mathbb{E}_{FCB}^{\min }(\mathbb{M})=0$. In Definition \ref{Def3.3}, the maximum of FCB entropy is defined as
$\mathbb{E}_{FCB}^{\max }(\mathbb{M})=\log ({{2}^{\left| \Theta  \right|}}-1)=\log ({{2}^{n}}-1)$. Therefore, the range of FCB entropy is $[0, \log ({{2}^{n}}-1)]$. So FCB entropy doesn’t satisfy Property \ref{pro.3}.
	
% \end{proo}
\end{pro}
%Property 4. Monotonicity
\begin{pro}[\textit{Monotonicity}]
\rm	Monotonicity means that in evidence theory, when information is significantly reduced (uncertainty increases), the measurement method of total uncertainty should not reduce the uncertainty.

Specifically, in DSET let two CBBAs ${{\mathbb{M}}_{1}}$ and ${{\mathbb{M}}_{2}}$ be defined on a FoD $\Theta =\{{{e}_{1}},{{e}_{2}},\cdots ,{{ e }_{n}}\}$. For any focal element ${{P}_{i}}\subseteq \Theta $ satisfying
\begin{equation}
% \small
	GBe{{l}_{1}}({{P}_{i}})\ge GBe{{l}_{2}}({{P}_{i}}),
\end{equation}
it must satisfy the following relationship:
\begin{equation}
% \small
	{{\mathbb{E}}_{FCB}}({{\mathbb{M}}_{1}})\le {{\mathbb{E}}_{FCB}}({{\mathbb{M}}_{2}}).
\end{equation}

\noindent \textbf{Proof 4.}
\rm	Negation plays a pivotal role in evidence theory, offering a valuable perspective for problem analysis. By examining events from a negative standpoint, it becomes possible to address issues that may not be resolvable from a positive perspective. As the count of negative iterations increases, evidence theory posits that the amount of information diminishes while uncertainty escalates. This paper employs the concept of negation to demonstrate the monotonicity of FCB entropy.

% After 10 iterations of negation, the evolution of CBBA in CET with respect to the number of negations is detailed in Table \ref{tab3}.

\noindent \textbf{Case 1.}
When CBBA is defined over the complex plane, FCB entropy is leveraged to gauge the uncertainty of CBBA, effectively handling complex-form information. The monotonicity is established through negative reasoning. In the context of CET, a novel exponential negation method for CBBA has been introduced \cite{yang2023exponential}, offering a technique to affirm the monotonicity of FCB entropy. Consider a Frame of Discernment (FoD) $\Theta =\{{{e}_{1}}, {{ e }_{2}}\}$ and select an initial value conducive to calculation.  The corresponding computations of FCB entropy are presented in Table \ref{tab:5}. Observations from Table \ref{tab:5} indicate that FCB entropy ascends with the increasing number of negations, thereby confirming the monotonicity of FCB entropy when applied to CBBA. However, it can be seen from Figure \ref{Fig.7} that QB entropy does not satisfy monotonicity.

\noindent \textbf{Case 2.}
According to Axiom \ref{axi3.1}, as a generalization of FB entropy, when CBBA degenerates to BBA, FCB entropy degenerates into FB entropy.
	Luo proposed a negative method of BBA. The negative direction is the direction of neglect. Similar to the process in \textbf{Case 1}. Consider a FoD $\Theta =\{{{e}_{1}}, {{ e }_{2}}\}$, select the initial value that is easy to calculate.  The change trend of FCB entropy and other previous entropies in evidence theroy with the number of iterations are shown in Figure \ref{Fig.7}. According to Figure \ref{Fig.7}, it’s clear to see that FB entropy \cite{zhou2022fractal} and FCB entropy increase with the number of iterations increasing, while Deng entropy \cite{Deng2020ScienceChina}, Pal et al.'s entropy \cite{pal1992uncertainty}, Zhou et al.'s entropy \cite{zhou2020weight} and Cui et al.'s entropy \cite{cui2019improved} first increase and then decrease with the number of iterations increasing. That is, FB entropy and FCB entropy meet the requirements of monotonicity, while Deng entropy, Pal et al.'s entropy, Zhou et al.'s entropy and Cui et al.'s entropy does not meet the requirements of monotonicity.
	\begin{figure}[H] %H为当前位置，!htb为忽略美学标准，htbp为浮动图形
	\centering %图片居中
	\includegraphics[width=0.5\textwidth]{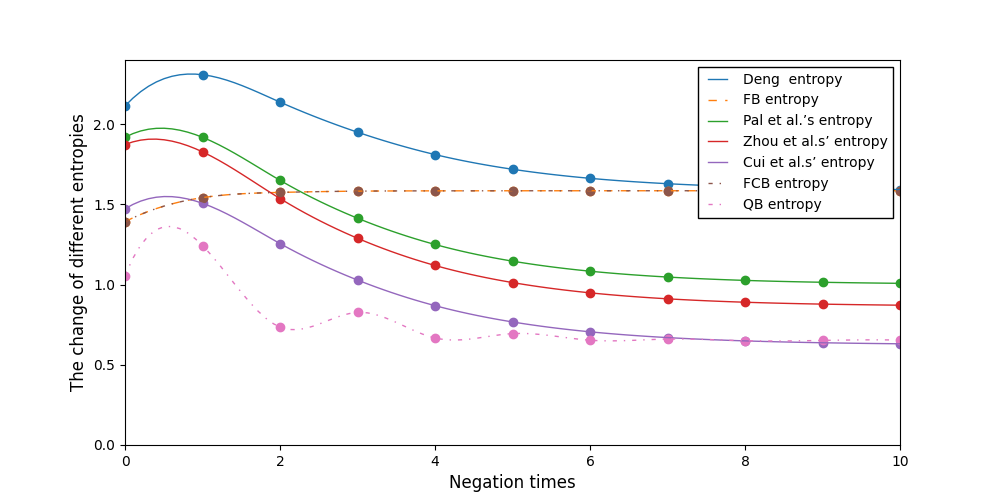} %插入图片，[]中设置图片大小，{}中是图片文件名
	\caption{The change trend of different entropies with the increase of negation times.} %最终文档中希望显示的图片标题
	\label{Fig.7} %用于文内引用的标签
\end{figure}	
% 	\begin{table}
% %		\tiny
% 	 \Huge
% 		\centering
%   %	\setlength{\tabcolsep}{15pt}
% 		\caption{Results of each iteration of different entropies.}\label{tab:4}%添加标题 设置标签
% 		\resizebox{.5\textwidth}{!}{
% 			\begin{threeparttable}
% 				\begin{tabular}{llllllllllll}
% 					\toprule
% 					& \multicolumn{11}{c}{Times} \\ \cmidrule(lr){2-12}
% 					Measures& 0& 1 &  2 & 3 & 4&5&6&7&8&9&10 \\
% 					\midrule
% 					Deng entropy \cite{Deng2020ScienceChina}&2.1133&2.3083&2.1375&1.9500&1.8105&1.7189&1.6624&1.6289&1.6095&1.5986&1.5924\\
% 					FB entropy \cite{zhou2022fractal}&1.3921&1.5420&1.5746&1.5824&1.5843&1.5848&1.5849&1.5850&1.5850&1.5850&1.5850\\
% 					Pal et al.s' entropy \cite{pal1992uncertainty}&1.9183&1.9182&1.6500&1.4138&1.2500&1.1461&1.0835&1.0470&1.0261&1.0144&1.0078\\
% 					Zhou et al.'s entropy \cite{zhou2020weight}&1.8728&1.8274&1.5364&1.2888&1.1192&1.0126&0.9486&0.9113&0.8900&0.8782&0.8715\\
% 					Cui et al.s' entropy \cite{cui2019improved} &1.4721&1.5068&1.2558&1.0283&0.8687&0.7671&0.7056&0.6696&0.6490&0.6374&0.6309\\
% 				FCB entropy&1.3921&1.5420&1.5746&1.5824&1.5843&1.5848&1.5849&1.5850&1.5850&1.5850&1.5850\\
% 					\bottomrule
% 				\end{tabular}
% 			\end{threeparttable}
% 		}
% 	\end{table}
% \end{proo}	
Combining the above two cases, it can be verified that FCB entropy satisfies monotonicity.
\end{pro}
%Property 5. Additivity
\begin{pro}[\textit{Additivity}]
\rm	Given two independent FoD $\Gamma  $ and $\Upsilon $, two CBBAs are defined on the two FoDs, respectively. Let $\Psi =\Gamma \times \Upsilon $ be a joint FoD. Then the total uncertainty measurement should satisfy
	\begin{equation}
 % \small
	{{\mathbb{E}}_{FCB}}({{\mathbb{M}}_{\Psi }})={{\mathbb{E}}_{FCB}}({{\mathbb{M}}_{\Gamma}})+{{\mathbb{E}}_{FCB}}({{\mathbb{M}}_{\Upsilon }}).
	\end{equation}
For BBA, there are two different definitions of joint BBA depending on whether $m(\varnothing )=0$. In this paper, assuming that CBBA is normalized, the framework of joint CBBA is defined as $\Gamma \times \Upsilon =({{2}^{\left| \Gamma  \right|}}-1)({{2}^{\left| \Upsilon  \right|}}-1)$.

According to the above definition, the number of complex quality functions of $\Psi $ is less than its power set. For FCB entropy, the average distribution of multifocal elements to power sets should be changed to the number of subsets assigned to this frame.  Joint CBBA ${{\mathbb{M}}^{\Psi }}$ and joint FCBBA $\mathbb{M}_{F}^{\Psi }$ are defined by
\begin{equation}
% \small
	\begin{aligned}
	&{{\mathbb{M}}^{\Psi }}({{\psi }_{ij}})=\mathbb{M}({{\tau }_{i}})\times \mathbb{M}({{\gamma }_{j}}),\\
	&	{{\mathbb{M}}^{\Psi }}({{\psi }_{ij}}{{\psi }_{im}})=\mathbb{M}({{\tau }_{i}})\times \mathbb{M}({{\gamma }_{j}}{{\gamma }_{m}}),\\
	&{{\mathbb{M}}^{\Psi }}({{\psi }_{ij}}{{\psi }_{im}}{{\psi }_{nj}}{{\psi }_{nm}})=\mathbb{M}({{\tau }_{i}}{{\tau }_{n}})\times \mathbb{M}({{\gamma }_{j}}{{\gamma }_{m}}),\\
	&\forall {{A}_{i}}\in {{2}^{\left| \Psi  \right|}},\mathbb{M}_{F}^{\Psi }({{A}_{i}})={{\mathbb{M}}^{\Psi }}({{A}_{i}})+\sum\limits_{{{A}_{i}}\subseteq {{D}_{i}};{{B}_{i}}\times {{C}_{i}}={{D}_{i}}}{\frac{{{\mathbb{M}}^{\Psi }}({{D}_{i}})}{S\left( {{B}_{i}},{{C}_{i}} \right)}},
	\end{aligned}
\end{equation}
where $S\left( {{B}_{i}},{{C}_{i}} \right)=({{2}^{\left| {{B}_{i}} \right|}}-1)({{2}^{\left| {{C}_{i}} \right|}}-1)$.
%\begin{equation}
%\end{equation}
%\begin{equation}
%\end{equation}
%\begin{equation}
%\end{equation}
% \begin{proo}

\noindent \textbf{Proof 5.}
\rm	Given $\psi \in {{2}^{\left| \Psi  \right|}}$, $\tau \in {{2}^{\left| \Gamma  \right|}}$ and $\gamma \in {{2}^{\left| \Upsilon  \right|}}$, and the relationship of the three FoDs is $\Psi =\Gamma \times \Upsilon $. According to Definition \ref{def3.1}, the following equation is derived
	\begin{equation}
 % \small
		\begin{aligned}
		 \mathbb{M}_{F}^{\Psi }(\psi )
			% &={{\mathbb{M}}^{\Psi }}(\psi )+\sum\limits_{{{A}_{i}}\subseteq {{D}_{i}};{{B}_{i}}\times {{C}_{i}}={{D}_{i}}}{\frac{{{\mathbb{M}}^{\Psi }}({{D}_{i}})}{S\left( {{B}_{i}},{{C}_{i}} \right)}} \\ 
			& =\mathbb{M}(\tau )\times \mathbb{M}(\gamma )+\sum\limits_{{{A}_{i}}\subseteq {{D}_{i}};{{B}_{i}}\times {{C}_{i}}={{D}_{i}}}{\frac{\mathbb{M}({{B}_{i}})\times \mathbb{M}({{C}_{i}})}{S\left( {{B}_{i}},{{C}_{i}} \right)}} \\ 
			% & =\left( \mathbb{M}(\tau )+\sum\limits_{\tau \subseteq {{B}_{i}}}{\frac{\mathbb{M}({{B}_{i}})}{{S\left( {{B}_{i}},{0} \right) }}} \right )\times \left( \mathbb{M}(\gamma )+\sum\limits_{\gamma \subseteq {{C}_{i}}}{\frac{\mathbb{M}({{C}_{i}})}{{S\left( {0},{{C}_{i}} \right)}}} \right) \\ 
			& =\mathbb{M}_{F}^{\Gamma }(\tau )\times \mathbb{M}_{F}^{\Upsilon }(\gamma ) \\ 
		\end{aligned}
	\end{equation}
then calculate the modulus of $M_{F}^{\Psi }(\psi )$ to obtain the following equation
\begin{equation}
% \small
	\left| \mathbb{M}_{F}^{\Psi }(\psi ) \right|=\left| \mathbb{M}_{F}^{\Gamma }(\tau ) \right|\times \left| \mathbb{M}_{F}^{\Upsilon }(\gamma ) \right|,
\end{equation}
next, normalize the obtained FCBBA modulus
\begin{equation}
% \small
% \begin{aligned}
 \mathbb{C}o{{m}_{F}}(\psi )
 % &=\frac{\left| \mathbb{M}_{F}^{\Psi }(\psi ) \right|}{\sum\limits_{{{\psi }_{i}}\in {{2}^{\Psi }}}{\left| \mathbb{M}_{F}^{\Psi }({{\psi }_{i}}) \right|}}\\
 =\frac{\left| \mathbb{M}_{F}^{\Gamma }(\tau ) \right|\times \left| \mathbb{M}_{F}^{\Upsilon }(\gamma ) \right|}{\sum\limits_{{{\tau }_{i}}\in {{2}^{\Gamma }};{{\gamma }_{i}}\in {{2}^{\Upsilon }}}{\left| \mathbb{M}_{F}^{\Gamma }({{\tau }_{i}}) \right|\times \left| \mathbb{M}_{F}^{\Upsilon }({{\gamma }_{i}}) \right|}}. \\ 
	 % \\ 
% \end{aligned}
\end{equation}

It is proved that the joint CBBA and FCBBA are continuous in the framework. In addition, it is also known that Shannon entropy is additive, and FCB entropy is similar to Shannon entropy. It is easy to prove that FCB entropy is additive. An example \ref{exa4.1} is given in appendix to verify that FCB entropy satisfies additivity.
% \end{proo}
\end{pro}

%Property 6. Subadditivity
\begin{pro}[\textit{Subadditivity}]
\rm
		Given two independent FoDs $\Gamma $ and $\Upsilon $, two CBBAs are defined on the two FoDs, respectively. Let $\Psi =\Gamma \times \Upsilon $ be a joint FoD. Then the total uncertainty measurement should satisfy
	 \begin{equation}
  % \small
{{\mathbb{E}}_{FCB}}({{\mathbb{M}}_{\Psi }})\le {{\mathbb{E}}_{FCB}}({{\mathbb{M}}_{\Gamma }})+{{\mathbb{E}}_{FCB}}({{\mathbb{M}}_{\Upsilon }}).
	 \end{equation}

\noindent \textbf{Proof 6.}
\rm	Whether the two FoDs $\Gamma $ and $\Upsilon $ are independent will have an impact on the entropy of the combined FoD $\Psi$, which is mainly divided into the following two cases.
	
\noindent \textbf{Case 1.} If CBBAs in $\Gamma $ and $\Upsilon$ are independent of each other, it can be known from Property 5 that ${{\mathbb{E}}_{FCB}}({{\mathbb{M}}_{\Psi }})={{\mathbb{E}}_{FCB}}({{\mathbb{M}}_{\Gamma }})+{{\mathbb{E}}_{FCB}}({{\mathbb{M}}_{\Upsilon }})$.

\noindent \textbf{Case 2.} If CBBAs in $\Gamma$ and $\Upsilon$ are not independent of each other, when the two CBBAs are jointed, there will be some overlapping information, resulting in reduced uncertainty, namely 
${{\mathbb{E}}_{FCB}}({{\mathbb{M}}_{\Psi }})<{{\mathbb{E}}_{FCB}}({{\mathbb{M}}_{\Gamma }})+{{\mathbb{E}}_{FCB}}({{\mathbb{M}}_{\Upsilon }})$.

Combining the above two cases, it can be concluded that FCB entropy satisfies the subadditivity.

\end{pro}

%Property 7.Calculation complexity
\begin{pro}[\textit{Time complexity}]

\rm	Let a CBBA $\mathbb{M}$ be defined on a FoD $\Theta =\{{{e}_{1}},{{e}_{2}},\cdots ,{{ e }_{n}}\}$ which contains $n$ elements. The measurement of FCB entropy can be divided into four steps. The total time complexity of each step and the algorithm is summarized in Table \ref{tab:1}.
	\begin{table}
	\centering
	% \resizebox{\textwidth}{!}{
		\begin{threeparttable}[b]
			\setlength{\tabcolsep}{6pt}
			\caption{Time complexity of each step in the propoed FCB entropy}\label{tab:1}%添加标题 设置标签
			\begin{tabular}{lll}
				\toprule
				Steps& Time complexity& Description\\
				\midrule
				Step 1& $O(n{{2}^{n}})$ & Calculate FCBBA based on CBBA\\
				Step 2& $O({{2}^{n}})$ & Calculate the modulus of FCBBA.\\
				Step 3& $O({{2}^{n}})$   &  Calculate $\mathbb{C}o{{m}_{F}}$ of each focal element.\\
				Step 4& $O({{2}^{n}})$  & Finally, calculate FCB entropy.\\
				Steps 1-4&$O(n{{2}^{n}})$&Time complexity.\\
				\bottomrule
			\end{tabular}
			%\caption{这是一张三线表}\label{tab:aStrangeTable}  标题放在这里也是可以的
			\begin{tablenotes}
				\footnotesize
				\item[] Explanation of notations in the table:
				\item[1] $n$: The number of baic event in the frame of discernment  $\Theta $.
				\item[2] $O( f\left( n \right) )$: The maximum time complexity.
			\end{tablenotes}
		\end{threeparttable}
	% }
\end{table}

% \noindent \textbf{Step 1:} First, calculate FCBBA according to the given initial value of CBBA and the rules in Definition \ref{def3.2}. The worst time complexity of this step is $O(n{{2}^{n}})$.

% \noindent \textbf{Step 2:} Calculate the modulus of the generated FCBBA. The time complexity of this step is $O({{2}^{n}})$.

% \noindent \textbf{Step 3:} Use the calculated modulus of FCBBA to calculate the $\mathbb{C}o{{m}_{F}}$ of each focal element according to the definition of $\mathbb{C}o{{m}_{F}}$. The time complexity of this step is $O({{2}^{n}})$.

% \noindent \textbf{Step 4:} Finally, according to the definition of FCB entropy, the time complexity of this step is $O({{2}^{n}})$.

According to Table \ref{tab:1}, it can be seen that the computational complexity of FCB entropy is similar to that of FB entropy \cite{zhou2022fractal} and JS entropy \cite{jirouvsek2018new}, and the computational complexity of some methods is higher than that of FCB entropy. Therefore, the computational complexity of FCB entropy is still within an acceptable range.
\end{pro}

%Property 8. Discord and non-specificity
\begin{pro}[\textit{Discord and non-specificity}]\label{pro.8}
	\rm	For a total uncertain measurement method, it should be able to be divided into two parts to measure discord and non-specificity.
\begin{definition}\label{def4.1}
	(FCB entropy’s discord).
\rm	Given a FoD $\Theta =\{{{e}_{1}},{{e}_{2}},\cdots ,{{ e }_{n}}\}$ and a CBBA $\mathbb{M}$ is defined in it, then the FCB entropy’s discord part is defined by
	\begin{equation}
 % \small
	\mathbb{E}_{FCB}^{\mathcal{D}}(\mathbb{M})=-\sum\limits_{i=1}^{n}{\mathbb{C}omCBet({{e}_{i}})\log \left( \mathbb{C}omCBet({{e}_{i}}) \right)},
	\end{equation}
in which
\begin{equation}
% \small
	\mathbb{C}omCBet({{e}_{i}})=\frac{\left| CBet({{e}_{i}}) \right|}{\sum\limits_{i=1}^{n}{\left| CBet({{e}_{i}}) \right|}},
\end{equation}
where $\sum\limits_{i=1}^{n}{\left| CBet({{e}_{i}}) \right|}$ is used to normalize $\left| CBet({{e}_{i}}) \right|$ and ensures the range of $\mathbb{C}omCBet({{e}_{i}})$ is in $[0,1]$.
\end{definition}
\begin{definition}(FCB entropy’s non-specificity).
	\rm Given a FoD $\Theta =\{{{e}_{1}}$, ${{e}_{2}}$, $\cdots$,${{ e }_{n}}\}$ and a CBBA $\mathbb{M}$ is defined in it. Then the FCB entropy’s non-specificity part is defined by
	\begin{equation}
 % \small
		\begin{aligned}
	\mathbb{E}_{FCB}^{\mathcal{N}}(\mathbb{M})=
	&{{\mathbb{E}}_{FCB}}(\mathbb{M})-\mathbb{E}_{FCB}^{\mathcal{D}}(\mathbb{M})\\=
	&\sum\limits_{i=1}^{n}{\mathbb{C}omCBet({{e}_{i}})\log \left( \mathbb{C}omCBet({{e}_{i}}) \right)}\\
	&-\sum\limits_{{{A}_{i}}\in {{2}^{\Theta }}}{\mathbb{C}o{{m}_{F}}({{A}_{i}})\log (\mathbb{C}o{{m}_{F}}({{A}_{i}}))}	
		\end{aligned}
	\end{equation}
\end{definition}

Among the existing entropy measures, some can explicitly decompose discord and non-specificity directly from their formulas, such as Deng entropy and JS entropy. However, akin to FB entropy, FCB entropy does not straightforwardly yield measures of discord and non-specificity from its formula. The FCB entropy's genesis lies in the Complex Pignistic Belief Transformation (CPBT) process. When CBBA is translated into probability via CPBT, the resulting uncertainty is solely attributed to discord within the framework. Hence, the definitions of discord and non-specificity in FCB entropy are as previously stated.

In Definition \ref{def4.1}, $\mathbb{E}_{FCB}^{\mathcal{D}}(\mathbb{M})$ signifies the uncertainty due to singletons, excluding the uncertainty stemming from set allocation in evidence theory. This makes it a fitting representation for discord in FCB entropy. The total uncertainty minus the discord-induced uncertainty leaves the uncertainty due to non-specificity. Consequently, FCB entropy adheres to Property \ref{pro.8}.

An example is provided to illustrate the interplay between discord and non-specificity within FCB entropy.

% \begin{figure}[htbp]
% 	\centering
% 	\subfloat[Discord change of FCB entropy.]{\includegraphics[width=.45\columnwidth]{Fig.8.png}}\hspace{5pt}
% 	\subfloat[Non-specificity change of FCB entropy.]{\includegraphics[width=.45\columnwidth]{Fig.9.png}}\\
% 	\subfloat[Title]{\includegraphics[width=.45\columnwidth]{Fig.10.png}}\hspace{5pt}
% 	\subfloat[Title]{\includegraphics[width=.45\columnwidth]{Fig.11.png}}
% 	\caption{Description.}
% \end{figure}
\begin{figure}[H] %H为当前位置，!htb为忽略美学标准，htbp为浮动图形
	\centering %图片居中
	\includegraphics[width=0.3\textwidth]{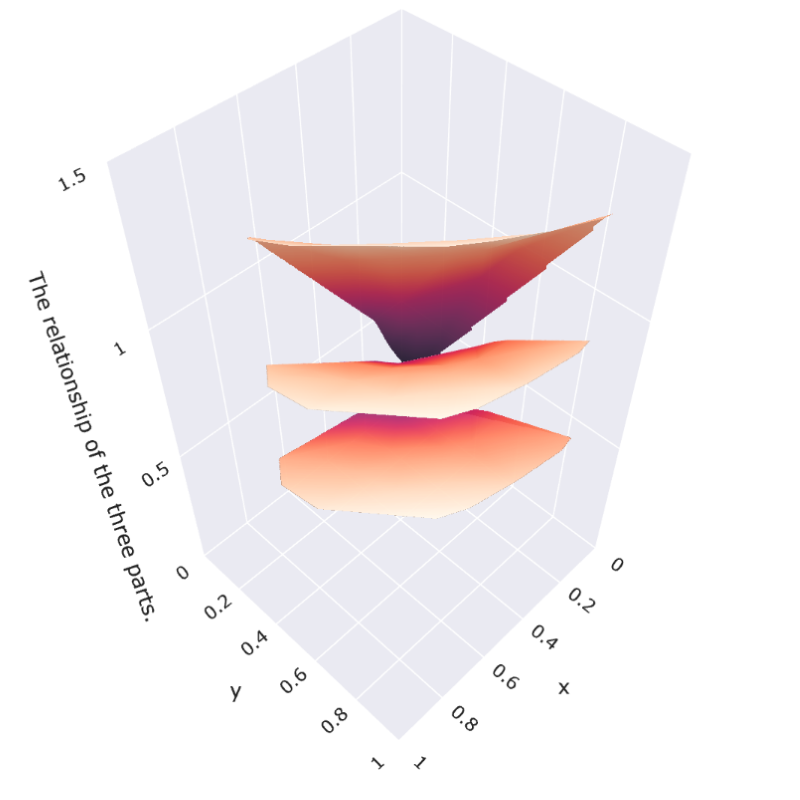} %插入图片，[]中设置图片大小，{}中是图片文件名
	\caption{The relationship between the three parts.} %最终文档中希望显示的图片标题
	\label{fig.11} %用于文内引用的标签
\end{figure}	

% \begin{figure} %%跨栏图为：figure*而单栏图为 figure
%     % \tiny
% % 	\centering
% % 	\subfloat[Discord change of FCB entropy.]{ \includegraphics[width=1.4in]{Fig.8} \label{fig.8}}
% %  % \hspace{2pt}  % 添加水平间距
% % 	\subfloat[Non-specificity change of FCB entropy.]{ \includegraphics[width=1.4in]{Fig.9} \label{fig.9}}
	
% % 	\quad  %用于子图换行 即这个上面是并列两子图 这个下面是另一子图
% % \subfloat[The total change of FCB entropy.]{ \includegraphics[width=1.4in]{Fig.10.png} \label{fig.10}}
% % \hspace{2pt}  % 添加水平间距
%         \subfloat[The relationship between the three parts.]{ \includegraphics[width=1.4in]{Fig.11.png} \label{fig.11}}
        
% \caption{Surface graphs of discord, non-specificity and total FCB entropy with $x$ and $y$.}
% 	\label{Fig.4}
% \end{figure}  %%跨栏图为：figure*而单栏图为 figure

\begin{exa}
\rm	Given a FoD $\Theta =\{{{e}_{1}},{{e}_{2}}\}$, let a CBBA $\mathbb{M}$ be defined in the FoD. There are also two variable parameters $x$ and $y$ in the defined CBBA. The CBBA is given as below
\begin{displaymath}
	\begin{aligned}
		&\mathbb{M}({{e}_{1}})=(1-x)-yi,\\
		&	\mathbb{M}({{e}_{1}},{{e}_{2}})=x+yi,\\
	\end{aligned}
\end{displaymath}
satisfying $\sqrt{{{x}^{2}}+{{y}^{2}}}\in [0,1]$ and $\sqrt{{{(1-x)}^{2}}+{{(-y)}^{2}}}\in [0,1]$.

Then according to Definition \ref{def3.1}, the FCBBA can be obtained
\begin{displaymath}
	\begin{aligned}
	&{{\mathbb{M}}_{F}}({{e}_{1}})=\mathbb{M}({{e}_{1}})+\frac{\mathbb{M}({{e}_{1}},{{e}_{2}})}{3}=\frac{3-2x}{3}-\frac{2y}{3}i,\\
	&{{\mathbb{M}}_{F}}({{e}_{2}})=\frac{\mathbb{M}({{e}_{1}},{{e}_{2}})}{3}=\frac{x}{3}+\frac{y}{3}i,\\
	&	{{\mathbb{M}}_{F}}({{e}_{1}},{{e}_{2}})=\frac{\mathbb{M}({{e}_{1}},{{e}_{2}})}{3}=\frac{x}{3}+\frac{y}{3}i.
	\end{aligned}
\end{displaymath}

With the change of the unknown number $x$ and $y$, 
% the change of the discord part, the non-specificity part of the FCB entropy and the overall FCB entropy are shown in Figures \ref{fig.8}, \ref{fig.9} and \ref{fig.10}. 
the relationship between overall FCB entropy and discord and non-specificity is shown in Figure \ref{fig.11}.
\end{exa}
\end{pro}
%Property 9. Sensitivity to change
% \begin{pro}[\textit{Sensitivity to change}]
% \rm A total uncertainty measurement should be very sensitive to changes in CBBA. Shannon entropy has been proved to be a very sensitive method. The solution form of FCB entropy is similar to that of Shannon entropy. And the change of multi-element focal element will lead to the change of ${{\mathbb{M}}_{F}}({{A}_{i}})$ of its many subsets, so FCB entropy is a sensitive method. In addition, the sensitivity to CBBA changes can also be intuitively reflected in many examples in this paper.
% \end{pro}

\section{APPLICATION}\label{se5}
In this section, we initially employ illustrative example to highlight the benefits of FCB entropy. Subsequently, we apply the pattern recognition algorithm from \cite{pan2023complex} and the information fusion processing algorithm from \cite{wu2024novel}, substituting the entropy measure with FCB entropy for experimental validation. The outcomes underscore the effectiveness of the proposed FCB entropy.

\begin{figure}[H] %H为当前位置，!htb为忽略美学标准，htbp为浮动图形
	\centering %图片居中
	\includegraphics[width=0.4\textwidth]{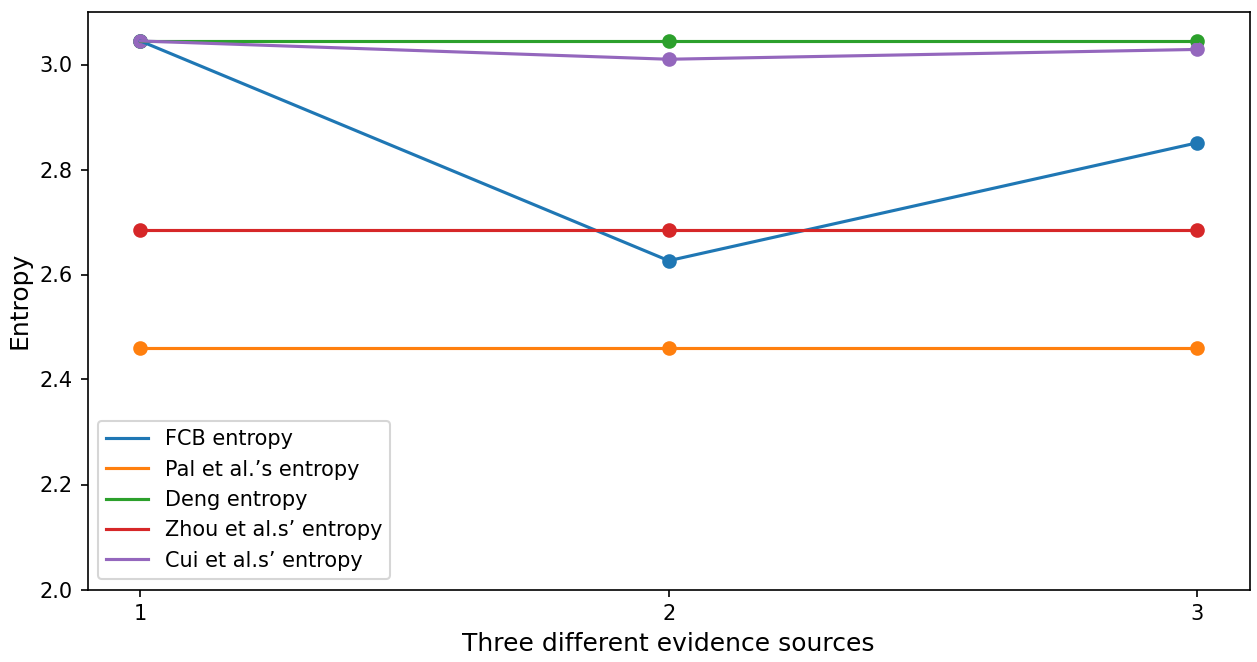} %插入图片，[]中设置图片大小，{}中是图片文件名
	\caption{Line chart of calculation results of different entropy in Example \ref{exa5.2}.} %最终文档中希望显示的图片标题
	\label{Fig.12} %用于文内引用的标签
\end{figure}	

\subsection{Discuss from numerical examples}
The construction of FCBBA entails the allocation of multi-element focal elements to subsets, capturing the effect of event intersections on the system's uncertainty assessment. Consequently, FCB entropy serves as an indicator of the uncertainty within CBBA that arises from overlapping events. To elucidate this concept, a toy example will be presented.

\begin{exa}\label{exa5.2}
 \rm In this example, there are three different CBBAs based on different FoDs, that is, the number of elements is different. The three CBBA values are as follows
\begin{displaymath}
	\begin{aligned}
	{{\mathbb{M}}_{1}}:
	&{{\mathbb{M}}_{1}}(\{{{x}_{1}},{{x}_{2}}\})=0.2+0.1i,{{\mathbb{M}}_{1}}(\{{{x}_{3}},{{x}_{4}}\})=0.6+0.2i,\\
	&{{\mathbb{M}}_{1}}(\{{{x}_{5}},{{x}_{6}}\})=0.2-0.3i,\\
	{{\mathbb{M}}_{2}}:
	&{{\mathbb{M}}_{2}}(\{{{x}_{1}},{{x}_{2}}\})=0.2+0.1i,{{\mathbb{M}}_{2}}(\{{{x}_{2}},{{x}_{3}}\})=0.6+0.2i,\\
	&{{\mathbb{M}}_{2}}(\{{{x}_{3}},{{x}_{6}}\})=0.2-0.3i,\\
	{{\mathbb{M}}_{3}}:
	&{{\mathbb{M}}_{3}}(\{{{x}_{1}},{{x}_{2}}\})=0.2+0.1i,{{\mathbb{M}}_{3}}(\{{{x}_{2}},{{x}_{3}}\})=0.6+0.2i,\\
	&{{\mathbb{M}}_{3}}(\{{{x}_{5}},{{x}_{6}}\})=0.2-0.3i.
	\end{aligned}
\end{displaymath}

Intuitively, the uncertainty of evidence source 1 should be the largest, and that of evidence source 2 should be the smallest. In order to compare the effect of FCB entropy and the previously proposed entropies on multi-element intersection, the previous entropies are extended to CBBA using Definition \ref{def2.16}. The calculation results of several different entropy of three different evidence sources are shown in Figure \ref{Fig.12}. From Figure \ref{Fig.12}, it can be seen that Pal et al.'s entropy, Deng entropy and Zhou et al.'s entropy cannot reflect the change in the amount of information caused by the intersection of focal elements. Cui et al.'s entropy and the proposed FCB entropy can reflect this feature, and the change trend of FCB entropy is more obvious than Cui et al.'s entropy, so FCB entropy is more
effective in reducing information loss and has a good description ability for uncertain information.

\end{exa}

\begin{figure*}[ht]
    \centering    
    \subfigure[The accuracy of classification results in the new Fractal-based complex belief entropy (FCBE), complex-valued Deng entropy (CDE) and quantum belief entropy (QBE) for Breast cancer data sets.]{				% 图片1([]内为子图标题)				
    \includegraphics[width=0.47\textwidth]{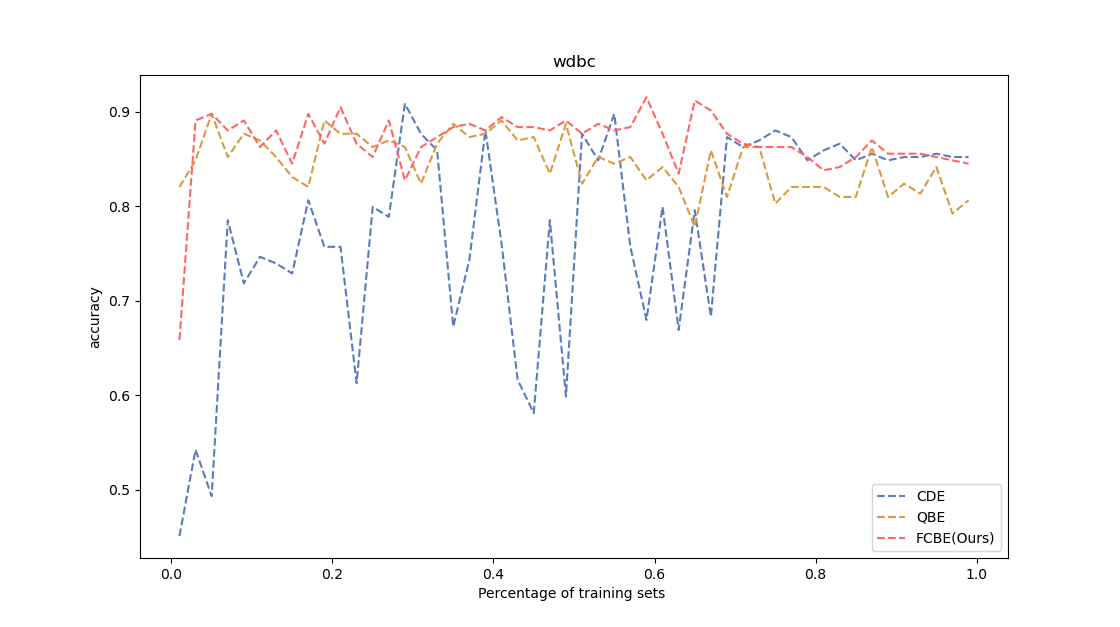}}\label{fig:a}			  % 子图1的图片宽度 不能空行
    \subfigure[The accuracy of classification results in the new Fractal-based complex belief entropy (FCBE) , complex-valued Deng entropy (CDE) and quantum belief entropy (QBE) for Glass data sets.]{				% 图片2
    \includegraphics[width=0.47\textwidth]{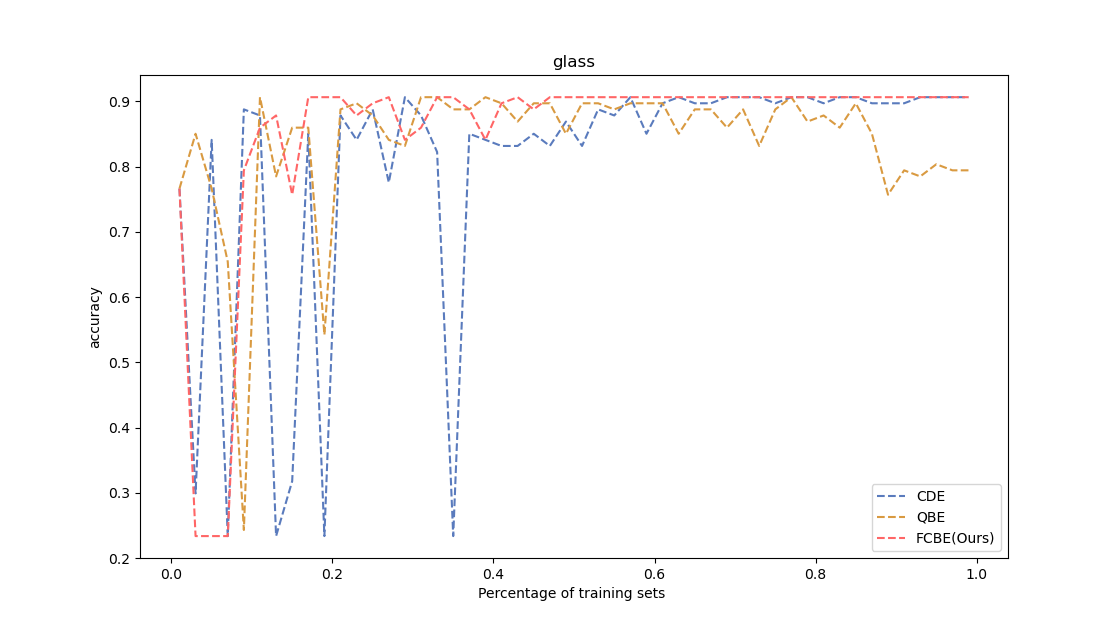}}\label{fig:b}
	% \caption{Data Screening} % 图片标题 
 
	\caption{The accuracy of classification results in Breast cancer and Glass data sets for FCBE and CDE.} % 图片标题 
    \label{fig:cl}
    % \vskip -0.3in
\end{figure*}

\subsection{Application in pattern classification}
In \cite{pan2023complex}, a pattern classification approach is presented on the basis of entropy. This section employs this approach to evaluate the novel FCB entropy. The method is divided into three stages:

\begin{itemize}
\item The initial stage involves securing the CBBAs for each piece of data in the test set. Compute the differences in the mean and standard deviation between the original training set and the one augmented with test data. According to the definition in \cite{pan2023complex}, the phase angle is represented by the mean value, while the modulus length corresponds to the standard deviation. The initial CBBA is derived as follows:
\begin{equation}
    \mathbb{M}\left( A \right)=\left( \frac{1}{{{e}^{\left| \Delta \delta  \right|}}} \right){{e}^{i\left( \left| \Delta \mu  \right| \right)}},
\end{equation}
where $\Delta \delta$ is the change value of the standard deviation, and $\Delta \mu$ is the change value of the mean. If the test set has $N$ data, $N$ sets of initial CBBAs are obtained. Then, the complex mass functions are normalized to ensure that each group satisfies:
\begin{equation}
    \sum{\mathbb{M}\left( A \right)}=1.
\end{equation}

\item The second stage is to determine the CBBA for the training set. Assuming the training set data comprises $n$ distinct classes, $M$ complex-valued masses must be obtained. For a given category, similar to the first stage, calculate the CBBAs for each data point. Subsequently, calculate the sum of differences in FCBE for each data point relative to all others. Entropy signifies the level of confusion; a smaller change in entropy indicates a better fit for the data set. Thus, the complex mass with the minimal entropy change is identified as the optimal mass for the training set.

\item The final stage classifies the test set data. Optimal CBBAs for the $n$ classes are obtained from the second stage. Mirroring the second stage's approach, a test set data point is selected, and the FCBE change value is determined using the complex mass function for each class. The smaller the entropy change, the more closely the test data aligns with the respective category. In other words, the test data is assigned to the category with the smallest entropy difference.
\end{itemize}

We utilized two classic datasets from the UCI repository (\href{http://archive.ics.uci.edu/ml/index.php}{http://archive.ics.uci.edu/ml/index.php}) for pattern classification: Breast Cancer and Glass. The classification outcomes are depicted in Figure \ref{fig:cl}. The x-axis represents the training set proportion, while the y-axis denotes classification accuracy. The test set remains constant, with the training ratio varying from 0.01 to 0.99 across 99 data points. Figure \ref{fig:cl}(a) illustrates the classification accuracy of the new FCB entropy (FCBE), Deng's complex-valued Deng entropy (CDE) and quantum belief entropy (QBE) on the Breast Cancer dataset. Figure \ref{fig:cl}(b) pertains to the Glass dataset. It is evident that the new FCB entropy outperforms CDE and QBE in terms of classification accuracy on both datasets.

Deng's experiment has established that complex entropy yields superior classification results compared to real entropy. Consequently, this paper refrains from conducting repetitive experiments. The enhancement is attributed to the fact that real entropy considers only the first element's information, directly related aspects such as length, density, and quality. In contrast, complex entropy incorporates the second element's information, indirectly related details. The phase angle encompasses this indirect information. As demonstrated in the experiment, the phase angle retains the mean's information. Even if the standard deviation is identical, differing means should classify two data points differently. Therefore, complex entropy accounts for the uncertainty of the second element, rendering it more effective than real entropy. Furthermore, in contrast to Deng’s CDE and Wu's QBE, the FCBE takes into account the influence of event intersections. In other words, the novel FCB entropy harnesses a greater wealth of information concealed within the intersections of distinct focal elements.

\begin{table*}
	\centering
	\caption{The modulus of different diseases' CBBAs in the fusion evidence.}\label{tab:7}%添加标题 设置标签
	% \resizebox{0.5\textwidth}{!}{
		\begin{threeparttable}
			\begin{tabular}{lcccccc}
				\toprule
				\multirow{3}{*}{Evidence fusion } &\multicolumn{3}{c}{Diseases} & \multirow{3}{*}{Evaluation results } & \multicolumn{2}{c}{Entropies} \\ 
                \cmidrule(lr){2-4}
				& Viral fever & Malaria & Typhoid & & QB entropy\cite{wu2024novel} & FCB entropy\\
				& $\left| \mathbb{M}(\{{{T}_{1}}\}) \right|$ & $\left| \mathbb{M}(\{{{T}_{2}}\}) \right|$&$\left| \mathbb{M}(\{{{T}_{3}}\}) \right|$ & \\
				\midrule
				${\mathbb{M}_{{{D}_{1}}}}\oplus {\mathbb{M}_{{{D}_{2}}}}$&$0.4105$&$0.4105$&$0.1086$&Cannot be determined & 2.0791 & \textbf{2.0636}\\
				${\mathbb{M}_{{{D}_{1}}}}\oplus {\mathbb{M}_{{{D}_{2}}}}\oplus {\mathbb{M}_{{{D}_{3}}}}$&$0.5248$&$0.3083$&$0.1648$&Viral fever & 2.0833 & \textbf{1.7369}\\
				${\mathbb{M}_{{{D}_{1}}}}\oplus {\mathbb{M}_{{{D}_{2}}}}\oplus {\mathbb{M}_{{{D}_{3}}}}\oplus {\mathbb{M}_{{{D}_{4}}}}$&$0.6733$&$0.2541$&$0.1096$&Viral fever & 1.8243 & \textbf{1.4138} \\
				${\mathbb{M}_{{{D}_{1}}}}\oplus {\mathbb{M}_{{{D}_{2}}}}\oplus {\mathbb{M}_{{{D}_{3}}}}\oplus {\mathbb{M}_{{{D}_{4}}}}\oplus {\mathbb{M}_{{{D}_{5}}}}$&$0.8432$&$0.1494$&$0.0766$&Viral frver & 1.3716 & \textbf{1.0286} \\
				\bottomrule
			\end{tabular}
		\end{threeparttable}
	% }
\end{table*}
\subsection{Application in information fusion}
We adopt the decision-making method proposed in the information fusion process in \cite{wu2024novel}, and conduct experimental comparisons between QB entropy and FCB entropy. The specific problem description and algorithm  are as follows:

% In real-world scenarios, the challenge of adjudicating a problem often arises due to an overwhelming array of evidence sources. To address this, the establishment of a threshold criterion becomes a strategic necessity. During the evidence fusion process, if the FCB entropy of the amalgamated evidence falls below this predetermined threshold, the resulting judgment can be deemed credible. This methodological approach not only maintains the integrity of the decision's accuracy but also enhances the overall efficiency of the decision-making process.

% In practical applications, when there are too many evidence sources for judging a problem, a threshold can be set. In the fusion process, when the FCB entropy of the fused evidence source is less than the set value, the judgment result can be considered credible. This algorithm can improve the decision-making efficiency on the premise of ensuring the correctness of a judgment.

\noindent\textbf{Problem statement:} Given a frame of discernment $\Theta =\{{{T}_{1}}$, $\ldots$, ${{T}_{i}}$, $\ldots$, ${{T}_{n}}\}$ and $N$ distinct evidence resources $\mathbb{M}=\{{{\mathbb{M}}_{1}}$, $\ldots$, ${{\mathbb{M}}_{j}}$, $\ldots {{\mathbb{M}}_{N}}\}$, the objective is to identify the target from $\Theta$ by integrating the minimal number of evidence sources necessary.
% The goal of the
% decision-making problem is to recognize the target from $\Theta =\{{{T}_{1}}$, $\ldots $, ${{T}_{i}}$, $\ldots$, ${{T}_{n}}\}$ by integrating as few evidence sources as possible.

The decision-making process is outlined as follows:

\noindent Step 1:
For each piece of evidences $j$, fusion results are computed using a complex evidence combination rule:
\begin{equation}
	Fus(\mathbb{M})={{({{({{\mathbb{M}}_{1}}\oplus {{\mathbb{M}}_{2}})}_{1}}\oplus \cdots \oplus {{\mathbb{M}}_{j}})}_{j-1}}.
\end{equation}

\noindent Step 2: 
A threshold value $\sigma $ is predefined for decision-making purposes.
% Assume there is a threshold value $\sigma $ which is set for the decision making.

If $\left| \mathbb{M}({{T}_{\delta }}) \right|< \sigma $, the target remains undetermined, and further fusion is required.

If $\left| \mathbb{M}({{T}_{\delta }}) \right|\ge \sigma $, then ${{T}_{\delta }}$ is considered a potential target:
\begin{equation}
	\begin{aligned}
		& \delta =\underset{1\le i\le n}{\mathop{\arg \max }}\,\{\left| \mathbb{M}(\{{{T}_{i}}\}) \right|\}, \\
		& \text{The potential target}\leftarrow {{T}_{\delta }}. \\
	\end{aligned}
\end{equation}

Subsequently, FCB entropy is computed using Eq. (\ref{eq.31}), to obtain ${{\mathbb{E}}_{FCB}}(Fus(\mathbb{M}))$.

A second threshold $\varepsilon $ is set to determine the conclusion of the fusion process.

If ${{\mathbb{E}}_{FCB}}(Fus(\mathbb{M}))\le \varepsilon $, the judgment is deemed credible, and the fusion process can terminate.

If ${{\mathbb{E}}_{FCB}}(Fus(\mathbb{M}))>\varepsilon $, the judgment remains indeterminate, necessitating further fusion.

%\noindent Step 2: Through calculating QB entropy by Eq. (\ref{eq35}), ${{E}_{Q}}(Fus(\mathbb{M}))$ can be obtained.
%
%\noindent Step 3: Assume there is a threshold value $\varepsilon $ which is set for the end of the fusion process.
%
%If ${{E}_{Q}}(Fus(\mathbb{M}))\le \varepsilon $, the judgment in the fusion result can be considered credible.
%
%If ${{E}_{Q}}(Fus(\mathbb{M}))>\varepsilon $, the judgment can not be determined. The fusion process needs to be continued.

\noindent Step 3: Repeat Steps 1-2 until the stopping criteria for the fusion process are met.

\begin{figure}[H] %H为当前位置，!htb为忽略美学标准，htbp为浮动图形
	\centering %图片居中 %用于文内引用的标签
	\includegraphics[width=0.5\textwidth]{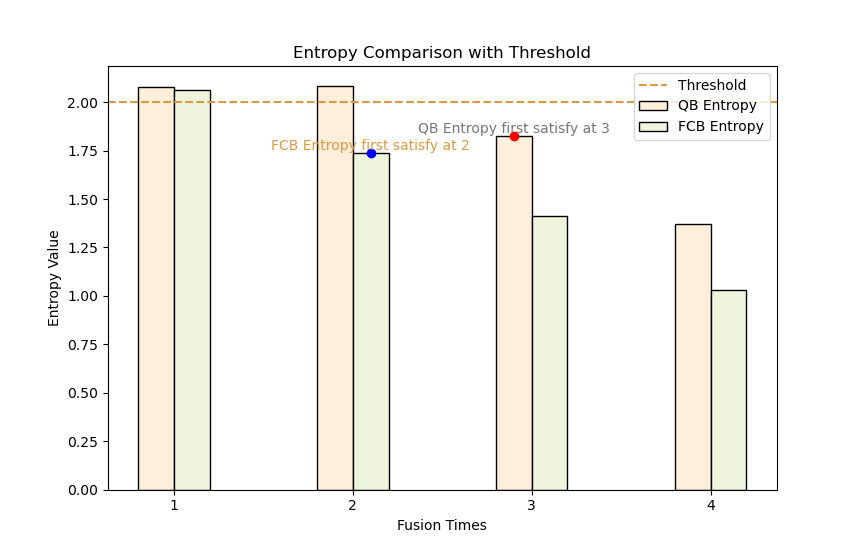} %插入图片，[]中设置图片大小，{}中是图片文件名
	\caption{Comparison of the two entropies in decision-making.} %最终文档中希望显示的图片标题
 \label{fig:com}
\end{figure}

We explore the practicality of FCB entropy through its application in a medical diagnostic context. The example presented is grounded in the complex model, reflecting the intricate nature of real-world data and application scenarios. This application not only demonstrates the versatility of the algorithm but also underscores its suitability for tackling complex diagnostic problems in the medical field. Additionally, we conducted a comparison with QB entropy \cite{wu2024novel}, an emerging measure of uncertainty within the field of CET. This comparison further underscores the efficacy of the novel approach, bringing its advantages into sharper relief.
The results are in Table \ref{tab:7}.

There are conflicts among the five evidence sources. As the fusion process progresses, the result of target recognition becomes more reliable. When ${\mathbb{M}_{{{D}_{1}}}}$,  ${\mathbb{M}_{{{D}_{2}}}}$ and ${\mathbb{M}_{{{D}_{3}}}}$ are fused, according to Table \ref{tab:7}, it can be seen that $\left| \mathbb{M}(\{{{D}_{1}}\}) \right|=0.5248$, $\left| \mathbb{M}(\{{{D}_{2}}\}) \right|=0.3083$ and $\left| \mathbb{M}(\{{{D}_{3}}\}) \right|=0.1648$. In this case, the patient $P$ is most likely to suffer from Viral fever with the belief value of 0.5248, which satisfies the threshold value 0.5. In the meantime, ${{\mathbb{E}}_{FCB}}(Fus(\mathbb{M}))=1.7369$ satisfy the threshold value 2.0. The recognition result under this fusion condition can be considered reliable. Then the fusion process ends and the decision-making problem is solved.

However, when ${\mathbb{M}_{{{D}_{1}}}}$,  ${\mathbb{M}_{{{D}_{2}}}}$ and ${\mathbb{M}_{{{D}_{3}}}}$ are fused,  ${{E}_{Q}}{{(Fus(\mathbb{M}))}_{\text{avg}}}=2.0833$ don't satisfy the threshold value 2.0. The recognition result of the fusion is considered to be unreliable and needs to continue the fusion process.  That is to say, using FCB entropy can make correct decisions faster and improve efficiency than QB entropy. And in the subsequent fusion process, we can also see that FCB entropy is very sensitive to changes in the amount of information after fusion, indicating the effectiveness of FCB entropy in measuring the uncertainty of CBBA. The specific comparison is shown in Figure \ref{fig:com}. 

The greatest contribution of FCB entropy is that it fully considers the impact of multi-element focal elements on its subset of focal elements, so it can mine more information from the frame of discernment, making it more advantageous in pattern recognition, information fusion, etc.

\section{Conclusions}\label{se6}
Complex evidence theory, as an advancement of the D-S evidence theory, has been pivotal in handling uncertainty across numerous domains. Yet, the precise measurement of uncertainty within complex evidence theory continues to be a subject of active research. This paper introduces a novel approach to link complex basic belief assignment with probability through a fractal-inspired complex pignistic belief transformation. Building on this, we define a new metric, Fractal-based belief (FCB) entropy, designed to quantify the uncertainty inherent in complex basic belief assignment. By examining the discord and non-specificity components, FCB entropy offers a nuanced perspective on uncertainty assessment. The properties of FCB entropy are thoroughly analyzed, and its practical utility is validated through various illustrative examples and practical application in pattern classification and information fusion. Looking forward, the application scope of FCB entropy will be further investigated, with the aim of refining decision-making processes and enhancing the understanding of complex systems within complex evidence theory. Moreover, our future research will indeed consider the decision-making capabilities of CBBAs on the complex plane. We plan to explore the geometric and algebraic properties of CBBAs to better understand their decision-making potential. This includes investigating how the complex plane can be utilized to represent and process belief structures that are inherently multidimensional.
% \section{Conclusions}\label{se6}
% As a generalization of D-S evidence theory, complex evidence theory has played an important role in many fields. However, the measurement of uncertainty is still an open issue. Many studies have proposed different measurement methods, such as belief entropy, divergence, etc. As a new subject, fractal theory has attracted many scholars' interest. In order to link CBBA with probability, a CPBT method is proposed to convert CBBA into probability. However, CPBT only gives the result of conversion without specific process. Therefore, the relationship between CBBA and probability cannot be comprehensively understood. Inspired by fractal theory, this paper proposes a fractal generation process of CPBT. Then, based on the generation process of CPBT, a new basic belief assignment called FCBBA is defined, and on this basis, FCB entropy is proposed to measure the uncertainty of CBBA. The discord and non-specificity are also defined. Then the properties of FCB entropy are analyzed, and several examples are used to verify its effectiveness. In the future, the application of FCB entropy in pattern recognition, evidence fusion and other fields will be analyzed and realized.

\section*{Acknowledgments}
The authors would like to express the sincere appreciation to the editor and anonymous reviewers for their insightful comments, which greatly improve the quality of this paper. This research is supported by the National Natural Science Foundation of China (No. 62003280), Chongqing Talents: Exceptional Young Talents Project (No. cstc2022ycjh-bgzxm0070), and Chongqing Overseas Scholars Innovation Program (No. cx2022024).

\bibliography{re}

\begin{thebibliography}{10}

\bibitem{zhou2019survey}
ZhiJie Zhou, GuanYu Hu, ChangHua Hu, ChengLin Wen, and LeiLei Chang.
\newblock A survey of belief rule-base expert system.
\newblock {\em IEEE Transactions on Systems, Man, and Cybernetics: Systems}, 51(8):4944--4958, 2019.

\bibitem{wang2022expert}
Zhen Wang, Xiaoyue Jin, Tao Zhang, Jiahao Li, Dengxiu Yu, Kang~Hao Cheong, and CL~Philip Chen.
\newblock Expert system-based multiagent deep deterministic policy gradient for swarm robot decision making.
\newblock {\em IEEE Transactions on Cybernetics}, 54(3):1614--1624, 2024.

\bibitem{liu2022orientational}
Zhun-ga Liu, Yi-min Fu, Quan Pan, and Zuo-wei Zhang.
\newblock Orientational distribution learning with hierarchical spatial attention for open set recognition.
\newblock {\em IEEE Transactions on Pattern Analysis and Machine Intelligence}, 45(7):8757 -- 8772, 2022.

\bibitem{meng2023novel}
Debiao Meng, Shiyuan Yang, Ab{\'\i}lio~MP de~Jesus, and Shun-Peng Zhu.
\newblock A novel kriging-model-assisted reliability-based multidisciplinary design optimization strategy and its application in the offshore wind turbine tower.
\newblock {\em Renewable Energy}, 203:407--420, 2023.

\bibitem{liao2023asynchronous}
Huchang Liao, Xiaowan Jin, Zeshui Xu, and Enrique Herrera-Viedma.
\newblock An asynchronous large-scale group decision-making method with punishment of unstable opinions and its application in traffic noise-control technologies selection.
\newblock {\em IEEE Transactions on Fuzzy Systems}, 32(2):510--523, 2024.

\bibitem{wang2024complex}
Zhen Wang, Haojing Li, Xiaoyue Jin, Dengxiu Yu, Kang~Hao Cheong, and Xuelong Li.
\newblock Complex continuous action iterated dilemma with incremental dynamic model.
\newblock {\em IEEE Transactions on Systems, Man, and Cybernetics: Systems}, 54(4):2309--2319, 2024.

\bibitem{chen2024risk}
Xingyuan Chen and Yong Deng.
\newblock Evidential software risk assessment model on ordered frame of discernment.
\newblock {\em Expert Systems with Applications}, 250:123786, 2024.

\bibitem{zhu2022zacm}
Ruonan Zhu, Qing Liu, Chongru Huang, and Bingyi Kang.
\newblock {Z-ACM: An approximate calculation method of Z-numbers for large data sets based on kernel density estimation and its application in decision-making}.
\newblock {\em Information Sciences}, 610:440--471, 2022.

\bibitem{FrameworkDNT2023}
Xinyang Deng and Wen Jiang.
\newblock A framework for the fusion of non-exclusive and incomplete information on the basis of {D} number theory.
\newblock {\em Applied Intelligence}, 53:11861--11884, 2023.

\bibitem{Fujita2020Hypotheses}
Hamido Fujita, Angelo Gaeta, Vincenzo Loia, and Francesco Orciuoli.
\newblock Hypotheses analysis and assessment in counter-terrorism activities: a method based on {OWA} and fuzzy probabilistic rough sets.
\newblock {\em IEEE Transactions on Fuzzy Systems}, 28:831--845, 2020.

\bibitem{ye2021novel}
Jin Ye, Jianming Zhan, Weiping Ding, and Hamido Fujita.
\newblock A novel fuzzy rough set model with fuzzy neighborhood operators.
\newblock {\em Information Sciences}, 544:266--297, 2021.

\bibitem{deng2022RPS}
Yong Deng.
\newblock Random permutation set.
\newblock {\em International Journal of Computers Communications \& Control}, 17(1):4542, 2022.

\bibitem{dempster1967upper}
Arthur~P Dempster.
\newblock Upper and lower probabilities induced by a multivalued mapping.
\newblock {\em Annals of Mathematical Statistics}, 38:325--339, 1967.

\bibitem{shafer1976mathematical}
Glenn Shafer.
\newblock {\em A mathematical theory of evidence}, volume~42.
\newblock Princeton university press, 1976.

\bibitem{cao2022network}
Bo~Cao, Chenghai Li, Yafei Song, Yueyi Qin, and Chen Chen.
\newblock {Network intrusion detection model based on CNN and GRU}.
\newblock {\em Applied Sciences}, 12(9):4184, 2022.

\bibitem{fei2022optimization}
Liguo Fei and Yanqing Wang.
\newblock {An optimization model for rescuer assignments under an uncertain environment by using {\uppercase{D}}empster--{\uppercase{S}}hafer theory}.
\newblock {\em Knowledge-Based Systems}, 255:109680, 2022.

\bibitem{fei2024evidential}
Liguo Fei, Xiaoyu Liu, and Changping Zhang.
\newblock {An evidential linguistic ELECTRE method for selection of emergency shelter sites}.
\newblock {\em Artificial Intelligence Review}, 57(4):81, 2024.

\bibitem{garg2020multiattribute}
Harish Garg and Shyi-Ming Chen.
\newblock Multiattribute group decision making based on neutrality aggregation operators of q-rung orthopair fuzzy sets.
\newblock {\em Information Sciences}, 517:427--447, 2020.

\bibitem{zhou2023large}
Mi~Zhou, Ya-Qian Zheng, Yu-Wang Chen, Ba-Yi Cheng, Enrique Herrera-Viedma, and Jian Wu.
\newblock A large-scale group consensus reaching approach considering self-confidence with two-tuple linguistic trust/distrust relationship and its application in life cycle sustainability assessment.
\newblock {\em Information Fusion}, 94:181--199, 2023.

\bibitem{Xiao2022GEJS}
Fuyuan Xiao.
\newblock {GEJS: A generalized evidential divergence measure for multisource information fusion}.
\newblock {\em IEEE Transactions on Systems, Man, and Cybernetics - Systems}, page DOI: 10.1109/TSMC.2022.3211498, 2022.

\bibitem{huang2023higherR}
Yingcheng Huang, Fuyuan Xiao, Zehong Cao, and Chin-Teng Lin.
\newblock Higher order fractal belief {R{\'e}nyi} divergence with its applications in pattern classification.
\newblock {\em IEEE Transactions on Pattern Analysis and Machine Intelligence}, 45(12):14709--14726, 2023.

\bibitem{huang2023fractal}
Yingcheng Huang, Fuyuan Xiao, Zehong Cao, and Chin-Teng Lin.
\newblock Fractal belief {R{\'e}nyi} divergence with its applications in pattern classification.
\newblock {\em IEEE Transactions on Knowledge and Data Engineering}, page DOI: 10.1109/TKDE.2023.3342907, 2023.

\bibitem{zhang2024divergence}
Lang Zhang and Fuyuan Xiao.
\newblock Belief {R{\'e}nyi} divergence of divergence and its application in time series classification.
\newblock {\em IEEE Transactions on Knowledge and Data Engineering}, page DOI: 10.1109/TKDE.2024.3369719, 2024.

\bibitem{xu2020evidence}
Xiaobin Xu, Deqing Zhang, Yu~Bai, Leilei Chang, and Jianning Li.
\newblock Evidence reasoning rule-based classifier with uncertainty quantification.
\newblock {\em Information Sciences}, 516:192--204, 2020.

\bibitem{tang2019perturbation}
Shuai-Wen Tang, Zhi-Jie Zhou, Chang-Hua Hu, Jian-Bo Yang, and You Cao.
\newblock Perturbation analysis of evidential reasoning rule.
\newblock {\em IEEE Transactions on Systems, Man, and Cybernetics: Systems}, 51(8):4895--4910, 2019.

\bibitem{fu2022extended}
Chao Fu, Bingbing Hou, Min Xue, Leilei Chang, and Weiyong Liu.
\newblock Extended belief rule-based system with accurate rule weights and efficient rule activation for diagnosis of thyroid nodules.
\newblock {\em IEEE Transactions on Systems, Man, and Cybernetics: Systems}, 53(1):251--263, 2023.

\bibitem{xu2024fault}
Xiaobin Xu, Haohao Guo, Zhenjie Zhang, Pengfei Shi, Wenguang Huang, Xiaoding Li, and Georg Brunauer.
\newblock Fault diagnosis method via one vs rest evidence classifier considering imprecise feature samples.
\newblock {\em Applied Soft Computing}, page 111761, 2024.

\bibitem{cao2020interpretability}
You Cao, Zhijie Zhou, Changhua Hu, Wei He, and Shuaiwen Tang.
\newblock On the interpretability of belief rule-based expert systems.
\newblock {\em IEEE Transactions on Fuzzy Systems}, 29(11):3489--3503, 2020.

\bibitem{li2022reliability}
Yan-Feng Li, Hong-Zhong Huang, Jinhua Mi, Weiwen Peng, and Xiaomeng Han.
\newblock {Reliability analysis of multi-state systems with common cause failures based on Bayesian network and fuzzy probability}.
\newblock {\em Annals of Operations Research}, 311:195--209, 2022.

\bibitem{ni2021towards}
Lei Ni, Yu-wang Chen, and Oscar de~Brujin.
\newblock Towards understanding socially influenced vaccination decision making: An integrated model of multiple criteria belief modelling and social network analysis.
\newblock {\em European Journal of Operational Research}, 293(1):276--289, 2021.

\bibitem{xiong2021conflicting}
Leihui Xiong, Xiaoyan Su, and Hong Qian.
\newblock Conflicting evidence combination from the perspective of networks.
\newblock {\em Information Sciences}, 580:408--418, 2021.

\bibitem{chang2021transparent}
Leilei Chang, Limao Zhang, Chao Fu, and Yu-Wang Chen.
\newblock Transparent digital twin for output control using belief rule base.
\newblock {\em IEEE Transactions on Cybernetics}, page DOI: 10.1109/TCYB.2021.3063285, 2021.

\bibitem{liu2021new}
Zhun-Ga Liu, Guang-Hui Qiu, Shu-Yue Wang, Tian-Cheng Li, and Quan Pan.
\newblock A new belief-based bidirectional transfer classification method.
\newblock {\em IEEE Transactions on Cybernetics}, 52(8):8101--8113, 2021.

\bibitem{zhang2022bsc}
Zuo-Wei Zhang, Zhun-Ga Liu, Arnaud Martin, and Kuang Zhou.
\newblock {BSC: Belief shift clustering}.
\newblock {\em IEEE Transactions on Systems, Man, and Cybernetics: Systems}, 53(3):1748--1760, 2022.

\bibitem{QuantumBPA2023}
Xinyang Deng, Siyu Xue, and Wen Jiang.
\newblock A novel quantum model of mass function for uncertain information fusion.
\newblock {\em Information Fusion}, 89:619--631, 2023.

\bibitem{xiao2020generalization}
{Xiao, Fuyuan}.
\newblock Generalization of {\uppercase{d}}empster--{\uppercase{s}}hafer theory: A complex mass function.
\newblock {\em Applied Intelligence}, 50(10):3266--3275, 2020.

\bibitem{xiao2020generalized}
{Xiao, Fuyuan}.
\newblock Generalized belief function in complex evidence theory.
\newblock {\em Journal of Intelligent \& Fuzzy Systems}, 38(4):3665--3673, 2020.

\bibitem{zhang2024Gaussian}
Shengjia Zhang, Mingrui Yin, Fuyuan Xiao, Zehong Cao, and Danilo Pelusi.
\newblock A complex {Gaussian} fuzzy numbers-based multisource information fusion for pattern classification.
\newblock {\em IEEE Transactions on Fuzzy Systems}, page DOI: 10.1109/TFUZZ.2024.3352615, 2024.

\bibitem{yang2023exponential}
Chengxi Yang and Fuyuan Xiao.
\newblock An exponential negation of complex basic belief assignment in complex evidence theory.
\newblock {\em Information Sciences}, 622:1228--1251, 2023.

\bibitem{Xiao2022NQMF}
Fuyuan Xiao and Witold Pedrycz.
\newblock Negation of the quantum mass function for multisource quantum information fusion with its application to pattern classification.
\newblock {\em IEEE Transactions on Pattern Analysis and Machine Intelligence}, page DOI: 10.1109/TPAMI.2022.3167045, 2022.

\bibitem{xiao2023generalized}
Fuyuan Xiao.
\newblock Generalized quantum evidence theory.
\newblock {\em Applied Intelligence}, 53(11):14329--14344, 2023.

\bibitem{Xiao2023QuantumXentropy}
Fuyuan Xiao.
\newblock {Quantum X-entropy in generalized quantum evidence theory}.
\newblock {\em Information Sciences}, 643:119177, 2023.

\bibitem{wu2024novel}
Keming Wu and Fuyuan Xiao.
\newblock A novel quantum belief entropy for uncertainty measure in complex evidence theory.
\newblock {\em Information Sciences}, 652:119744, 2024.

\bibitem{he2024quantumrule}
Huaping He and Fuyuan Xiao.
\newblock {A novel quantum {\uppercase{D}}empster's rule of combination for pattern classification}.
\newblock {\em Information Sciences}, page DOI: 10.1016/j.ins.2024.120617, 2024.

\bibitem{LEI2022112136}
Mingli Lei and Kang~Hao Cheong.
\newblock Node influence ranking in complex networks: A local structure entropy approach.
\newblock {\em Chaos, Solitons \& Fractals}, 160:112136, 2022.

\bibitem{zhu2024fractal}
Li~Zhu, Qianli Zhou, Yong Deng, and Kang~Hao Cheong.
\newblock Fractal-based basic probability assignment: A transient mass function.
\newblock {\em Information Sciences}, 652:119767, 2024.

\bibitem{hohle1982entropy}
Ulrich Hohle.
\newblock Entropy with respect to plausibility measures.
\newblock In {\em Proc. of 12th IEEE Int. Symp. on Multiple Valued Logic, Paris, 1982}, 1982.

\bibitem{dubois1987properties}
Didier Dubois and Henri Prade.
\newblock Properties of measures of information in evidence and possibility theories.
\newblock {\em Fuzzy Sets and Systems}, 24(2):161--182, 1987.

\bibitem{pal1992uncertainty}
Nikhil~R Pal, James~C Bezdek, and Rohan Hemasinha.
\newblock Uncertainty measures for evidential reasoning i: A review.
\newblock {\em International Journal of Approximate Reasoning}, 7(3-4):165--183, 1992.

\bibitem{jousselme2006measuring}
A-L Jousselme, Chunsheng Liu, Dominic Grenier, and {\'E}loi Boss{\'e}.
\newblock Measuring ambiguity in the evidence theory.
\newblock {\em IEEE Transactions on Systems, Man, and Cybernetics-Part A: Systems and Humans}, 36(5):890--903, 2006.

\bibitem{Deng2020ScienceChina}
Yong Deng.
\newblock Uncertainty measure in evidence theory.
\newblock {\em Science China Information Sciences}, 63(11):210201, 2020.

\bibitem{jirouvsek2018new}
Radim Jirou{\v{s}}ek and Prakash~P Shenoy.
\newblock A new definition of entropy of belief functions in the {\uppercase{d}}empster--{\uppercase{s}}hafer theory.
\newblock {\em International Journal of Approximate Reasoning}, 92:49--65, 2018.

\bibitem{PlEntropy2023}
Yebi Cui and Xinyang Deng.
\newblock Plausibility entropy: a new total uncertainty measure in evidence theory based on plausibility function.
\newblock {\em IEEE Transactions on Systems, Man, and Cybernetics: Systems}, 53(6):3833--3844, 2023.

\bibitem{cao2019extraction}
Zehong Cao, Chin-Teng Lin, Kuan-Lin Lai, Li-Wei Ko, Jung-Tai King, Kwong-Kum Liao, Jong-Ling Fuh, and Shuu-Jiun Wang.
\newblock {Extraction of SSVEPs-based inherent fuzzy entropy using a wearable headband EEG in migraine patients}.
\newblock {\em IEEE Transactions on Fuzzy Systems}, 28(1):14--27, 2019.

\bibitem{liu2021double}
Peide Liu, Mengjiao Shen, Fei Teng, Baoying Zhu, Lili Rong, and Yushui Geng.
\newblock Double hierarchy hesitant fuzzy linguistic entropy-based {TODIM} approach using evidential theory.
\newblock {\em Information Sciences}, 547:223--243, 2021.

\bibitem{wang2018uncertainty}
Xiaodan Wang and Yafei Song.
\newblock Uncertainty measure in evidence theory with its applications.
\newblock {\em Applied Intelligence}, 48(7):1672--1688, 2018.

\bibitem{yang2016new}
Yi~Yang and Deqiang Han.
\newblock A new distance-based total uncertainty measure in the theory of belief functions.
\newblock {\em Knowledge-Based Systems}, 94:114--123, 2016.

\bibitem{han2018belief}
Deqiang Han, Jean Dezert, and Yi~Yang.
\newblock Belief interval-based distance measures in the theory of belief functions.
\newblock {\em IEEE Transactions on Systems, Man, and Cybernetics: Systems}, 48(6):833--850, 2018.

\bibitem{yager2008entropy}
{Yager, Ronald R}.
\newblock Entropy and specificity in a mathematical theory of evidence.
\newblock In {\em Classic works of the {\uppercase{D}}empster-{\uppercase{S}}hafer theory of belief functions}, pages 291--310. Springer, 2008.

\bibitem{pan2023complex}
Lipeng Pan and Yong Deng.
\newblock Complex-valued deng entropy.
\newblock {\em Applied Intelligence}, 53(18):21201--21210, 2023.

\bibitem{zhou2022fractal}
Qianli Zhou and Yong Deng.
\newblock Fractal-based belief entropy.
\newblock {\em Information Sciences}, 587:265--282, 2022.

\bibitem{wang2018exploiting}
Zhen Wang, Marko Jusup, Lei Shi, Joung-Hun Lee, Yoh Iwasa, and Stefano Boccaletti.
\newblock Exploiting a cognitive bias promotes cooperation in social dilemma experiments.
\newblock {\em Nature Communications}, 9(1):2954, 2018.

\bibitem{wang2017onymity}
Zhen Wang, Marko Jusup, Rui-Wu Wang, Lei Shi, Yoh Iwasa, Yamir Moreno, and J{\"u}rgen Kurths.
\newblock Onymity promotes cooperation in social dilemma experiments.
\newblock {\em Science Advances}, 3(3):e1601444, 2017.

\bibitem{Wang2020CommunicatingSA}
Zhen Wang, Marko Jusup, Hao Guo, Lei Shi, Sunana Geek, Madhur Anand, Matja Perc, Chris~T. Bauch, J{\"u}rgen Kurths, Stefano Boccaletti, and Hans~Joachim Schellnhuber.
\newblock Communicating sentiment and outlook reverses inaction against collective risks.
\newblock {\em Proceedings of the National Academy of Sciences of the United States of America}, 117:17650 -- 17655, 2020.

\bibitem{wang2022modelling}
Zhen Wang, Chunjiang Mu, Shuyue Hu, Chen Chu, and Xuelong Li.
\newblock Modelling the dynamics of regret minimization in large agent populations: a master equation approach.
\newblock In {\em 2022 the 31st International Joint Conference on Artificial Intelligence, (IJCAI-22)}, pages 534--540, 2022.

\bibitem{huang2023some}
Junjie Huang, Yi~Fan, and Fuyuan Xiao.
\newblock On some bridges to complex evidence theory.
\newblock {\em Engineering Applications of Artificial Intelligence}, 117:105605, 2023.

\bibitem{shannon1948mathematical}
Claude~Elwood Shannon.
\newblock A mathematical theory of communication.
\newblock {\em The Bell system technical journal}, 27(3):379--423, 1948.

\bibitem{higashi1982measures}
Masahiko Higashi and George~J Klir.
\newblock Measures of uncertainty and information based on possibility distributions.
\newblock {\em International journal of general systems}, 9(1):43--58, 1982.

\bibitem{zhou2020weight}
Mi~Zhou, Yu-Wang Chen, Xin-Bao Liu, Ba-Yi Cheng, and Jian-Bo Yang.
\newblock Weight assignment method for multiple attribute decision making with dissimilarity and conflict of belief distributions.
\newblock {\em Computers \& Industrial Engineering}, 147:106648, 2020.

\bibitem{cui2019improved}
Huizi Cui, Qing Liu, Jianfeng Zhang, and Bingyi Kang.
\newblock An improved deng entropy and its application in pattern recognition.
\newblock {\em IEEE Access}, 7:18284--18292, 2019.

\bibitem{chen2023novel}
Xingyuan Chen and Yong Deng.
\newblock A novel combination rule for conflict management in data fusion.
\newblock {\em Soft Computing}, 27(22):16483--16492, 2023.

\bibitem{zhao2024linearity}
Tong Zhao, Zhen Li, and Yong Deng.
\newblock Linearity in deng entropy.
\newblock {\em Chaos, Solitons \& Fractals}, 178:114388, 2024.

\bibitem{tang2023improved}
Yongchuan Tang, Shiting Tan, and Deyun Zhou.
\newblock An improved failure mode and effects analysis method using belief {\uppercase{j}}ensen--{\uppercase{s}}hannon divergence and entropy measure in the evidence theory.
\newblock {\em Arabian Journal for Science and Engineering}, 48(5):7163--7176, 2023.

\bibitem{kharazmi2023deng}
Omid Kharazmi and Javier~E Contreras-Reyes.
\newblock Deng--fisher information measure and its extensions: Application to conway’s game of life.
\newblock {\em Chaos, Solitons \& Fractals}, 174:113871, 2023.

\bibitem{ozkan2023new}
K{\"u}r{\c{s}}ad {\"O}zkan, Ahmet Mert, and Serkan {\"O}zdemir.
\newblock A new proposed glcm texture feature: modified {\uppercase{r}}{\'e}nyi {\uppercase{d}}eng entropy.
\newblock {\em The Journal of Supercomputing}, 79(18):21507--21527, 2023.

\bibitem{Qiang2022fractal}
Chenhui Qiang, Yong Deng, and Kang~Hao Cheong.
\newblock Information fractal dimension of mass function.
\newblock {\em Fractals}, 30:2250110, 2022.

\bibitem{abellan2008requirements}
Joaqu{\'\i}n Abell{\'a}n and Andr{\'e}s Masegosa.
\newblock Requirements for total uncertainty measures in {\uppercase{d}}empster--{\uppercase{s}}hafer theory of evidence.
\newblock {\em International journal of general systems}, 37(6):733--747, 2008.

\end{thebibliography}

\clearpage
\appendix
In appendix, we give corresponding proofs for Axiom \ref{axi3.1},  Definition \ref{Def3.3} in the paper.

\begin{proo}\label{pro3.1}
\rm	When CBBA degenerates to BBA, the following formula can be obtained:
\begin{equation}
% \small
	{{\mathbf{m}}_{F}}({{A}_{k}})=\frac{\mathbf{m}({{A}_{k}})}{{{2}^{\left| {{A}_{k}} \right|}}-1}+\sum\limits_{{{A}_{k}}\subseteq {{B}_{k}}\cap \left| {{A}_{k}} \right|<\left| {{B}_{k}} \right|}{\frac{\mathbf{m}({{B}_{k}})}{{{2}^{\left| {{B}_{k}} \right|}}-1}}.
\end{equation}
Because BBA is defined in the real number field, there is no interference effect, namely, $\sum\limits_{{{A}_{k}}\subseteq {{B}_{k}}\cap \left| {{A}_{k}} \right|<\left| {{B}_{k}} \right|}{IE(B_{k})=0}$.
\begin{equation}
% \small
\left| {{\mathbf{m}}_{F}}({{A}_{k}}) \right|={{\mathbf{m}}_{F}}({{A}_{k}}),
\end{equation}
% \begin{equation}
% \small
% 	\begin{aligned}
% 	\left| {{\mathbf{m}}_{F}}({{A}_{k}}) \right|
% 	&=\left| \frac{\mathbf{m}({{A}_{k}})}{{{2}^{\left| {{A}_{k}} \right|}}-1}+\sum\limits_{{{A}_{k}}\subseteq {{B}_{k}}\cap \left| {{A}_{k}} \right|<\left| {{B}_{k}} \right|}{\frac{\mathbf{m}({{B}_{k}})}{{{2}^{\left| {{B}_{k}} \right|}}-1}} \right|\\
% 	&=\sqrt{{{\sum\limits_{{{C}_{k}}\in {{B}_{k}}}{\left| \frac{\mathbb{M}({{C}_{k}})}{{{2}^{\left| {{C}_{k}} \right|}}-1} \right|}}^{2}}+\sum\limits_{{{A}_{k}}\subseteq {{B}_{k}}\cap \left| {{A}_{k}} \right|<\left| {{B}_{k}} \right|}{IE({{B}_{k}})}}\\
% 	&={{\mathbf{m}}_{F}}({{A}_{k}}),
% 	\end{aligned}
% \end{equation}
since the value range of BBA is [0,1],
\begin{equation}
% \small
	\mathbb{C}o{{m}_{F}}({{A}_{k}})=\frac{\left| {{\mathbf{m}}_{F}}({{A}_{k}}) \right|}{\sum\limits_{B\in \Theta }{\left| {{\mathbf{m}}_{F}}({{B}_{k}}) \right|}}={{\mathbf{m}}_{F}}({{A}_{k}}).
\end{equation}
Then simplify the FCB entropy to obtain
\begin{equation}
% \small
	\begin{aligned}
		{{\mathbb{E}}_{FCB}}(m)={{E}_{FB}}(m)
		&=-\sum\limits_{{{A}_{k}}\in {{2}^{\Theta }}}{\mathbb{C}o{{m}_{F}}({{A}_{k}})\log (\mathbb{C}o{{m}_{F}}({{A}_{k}}))}\\
		&=-\sum\limits_{{{A}_{k}}\in {{2}^{\Theta }}}{{{m}_{F}}({{A}_{k}})\log ({{m}_{F}}({{A}_{k}}))}.
	\end{aligned}
\end{equation}
\end{proo}
Therefore, FCB entropy can be regarded as the generalization of FB entropy, which has better ability to process complex information than FB entropy.

\begin{proo}\label{pro3.3}
	\rm First, FCB entropy is expressed as a function
	\begin{equation}
 % \small
	{{f}_{FCB}}({{A}_{1}}, \cdots ,{{A}_{{{2}^{\left| \Theta  \right|}}-1}})=-\sum\limits_{{{A}_{k}}\in {{2}^{\Theta }}}{\mathbb{C}o{{m}_{F}}({{A}_{k}})\log (\mathbb{C}o{{m}_{F}}({{A}_{k}}))},
	\end{equation} 
 according to (\ref{eq.34}), that is, $\sum\limits_{{{A}_{k}}\in {{2}^{\Theta }}}{\mathbb{C}o{{m}_{F}}({{A}_{k}})}=1$. Therefore, the Lagrange function can be constructed as
	\begin{equation}
 % \small
		\begin{aligned}
			{{\mathbb{F}}_{FCB}}({{A}_{1}},{{A}_{2}},\cdots ,{{A}_{{{2}^{\left| \Theta  \right|}}-1}},\lambda )&=
		{{f}_{FCB}}({{A}_{1}},{{A}_{2}},\cdots ,{{A}_{{{2}^{\left| \Theta  \right|}}-1}})\\
			&+\lambda (\sum\limits_{{{A}_{k}}\in {{2}^{\Theta }}}{\mathbb{C}o{{m}_{F}}({{A}_{k}})}-1),
		\end{aligned}
	\end{equation}
then calculate the first partial derivative of 
${{\mathbb{F}}_{FCB}}({{A}_{1}},{{A}_{2}},\cdots ,{{A}_{{{2}^{\left| \Theta  \right|}}-1}})$.
% ${{f}_{FCB}}({{A}_{1}},{{A}_{2}},\cdots ,{{A}_{{{2}^{\left| \Theta  \right|}}-1}})=-\sum\limits_{{{A}_{k}}\in {{2}^{\Theta }}}{\mathbb{C}o{{m}_{F}}({{A}_{k}})\log (\mathbb{C}o{{m}_{F}}({{A}_{k}}))}$,
% \begin{equation}
% \small
% \begin{aligned}
% \frac{\partial {{\mathbb{F}}_{FCB}}({{A}_{1}},\cdots ,{{A}_{{{2}^{\left| \Theta  \right|}}-1}},\lambda )}{\partial \mathbb{C}o{{m}_{F}}({{A}_{k}})}&=-\log (\mathbb{C}o{{m}_{F}}({{A}_{k}}))-\frac{1}{\ln a}+\lambda \\
%      &=0,
% \end{aligned}
% \end{equation}
For all $\mathbb{C}o{{m}_{F}}({{A}_{k}})$,
\begin{equation}
% \small
	\log (\mathbb{C}o{{m}_{F}}({{A}_{k}}))=\lambda -\frac{1}{\ln a}=k,
\end{equation}
then according to (\ref{eq.34}),
\begin{equation}
% \small
	\sum\limits_{{{A}_{k}}\in {{2}^{\Theta }}}{\mathbb{C}o{{m}_{F}}({{A}_{k}})}=({{2}^{\left| \Theta  \right|}}-1)\mathbb{C}o{{m}_{F}}({{A}_{k}})=1,
\end{equation}
so when $\mathbb{C}o{{m}_{F}}({{A}_{k}})=\frac{1}{({{2}^{\left| \Theta  \right|}}-1)}$, FCB entropy reaches its maximum,
\begin{equation}
% \small
\mathbb{E}_{CFB}^{\max }(\mathbb{M})=-\sum\limits_{{{A}_{k}}\in {{2}^{\Theta }}}{\frac{1}{({{2}^{\left| \Theta  \right|}}-1)}\log \frac{1}{({{2}^{\left| \Theta  \right|}}-1)}}=\log ({{2}^{\left| \Theta  \right|}}-1).
\end{equation}
\end{proo}

\begin{exa} \label{exa4.1}
\rm	Given two independent FoD $X=\{{{x}_{1}},{{x}_{2}}\}$ and $Y=\{{{y}_{1}},{{y}_{2}}\}$, two CBBAs are defined on the two FoDs, respectively. Let $Z=X\times Y$ be a joint FoD. And the complex mass functions are
\begin{displaymath}
	\begin{aligned}
{{\mathbb{M}}^{X}}:&{{\mathbb{M}}^{X}}({{x}_{1}})=0.2+0.1i,{{\mathbb{M}}^{X}}({{x}_{2}})=0.5+0.1i,\\
 &{{\mathbb{M}}^{X}}({{x}_{1}},{{x}_{2}})=0.3-0.2i,\\
 {{\mathbb{M}}^{Y}}:&{{\mathbb{M}}^{Y}}({{y}_{1}})=0.3+0.2i,{{\mathbb{M}}^{Y}}({{y}_{2}})=0.2+0.1i,\\
 &{{\mathbb{M}}^{Y}}({{y}_{1}},{{y}_{2}})=0.5-0.3i.\\
	\end{aligned}
\end{displaymath}

Using the normalized $\mathbb{C}o{{m}_{F}}$ to calculate the FCB entropy, it can be obtained that
\begin{displaymath}
	{{\mathbb{E}}_{FCB}}({{\mathbb{M}}_{Z}})=2.8317={{\mathbb{E}}_{FCB}}({{\mathbb{M}}_{X}})+{{\mathbb{E}}_{FCB}}({{\mathbb{M}}_{Y}}),
\end{displaymath}
which indicates that FCB entropy satisfies additivity.
	\end{exa}

\vfill

\end{document}